\documentclass[%
 reprint, superscriptaddress,
 amsmath,amssymb,
 aps, pre
]{revtex4-2}

\usepackage{graphicx}
\usepackage{dcolumn}
\usepackage{bm}
\usepackage{titlesec}
\usepackage{mathrsfs}

\begin{document}

\preprint{APS/123-QED}

\title{Diffusion of a single colloid on the surface of a giant vesicle and a droplet}

\author{Clément Marque}
 \homepage{cmarque@ics-cnrs.unistra.fr}
\affiliation{Institut Charles Sadron, CNRS UPR-22, 23 rue du Loess, Strasbourg, France
}%

\author{Gaetano D’Avino}
\affiliation{Dipartimento di Ingegneria Chimica, dei Materiali e della Produzione Industriale,
Università degli Studi di Napoli Federico II, P.le Tecchio 80, 80125 Napoli, Italy
}%

\author{Domenico Larobina}
\affiliation{Institute of Polymers, Composites and Biomaterials, National Research Council of
Italy, Napoli, 80055 Portici, Italy
}%

\author{Aude Michel}
\affiliation{Sorbonne Université, CNRS, PHysicochimie des Électrolytes et Nanosystèmes InterfaciauX (PHENIX), F-75005 Paris, France; Institut Universitaire de France (IUF), 75231 Paris, France}%

\author{Ali Abou-Hassan}
\affiliation{Sorbonne Université, CNRS, PHysicochimie des Électrolytes et Nanosystèmes InterfaciauX (PHENIX), F-75005 Paris, France; Institut Universitaire de France (IUF), 75231 Paris, France}%

\author{Antonio Stocco}
\homepage{stocco@unistra.fr}
\affiliation{Institut Charles Sadron, CNRS UPR-22, 23 rue du Loess, Strasbourg, France
}%

\date{\today}

\begin{abstract}

The study of interactions between biomimetic membranes and micron-sized particles is crucial for understanding various biological processes. Here, we control microparticle spontaneous engulfment by giant lipid vesicles by tuning particle surface charge, exploring regimes from negligible to strong adhesion. We focus our attention on dissipative phenomena at the micron- and nano- scales, occurring when a particle is wrapped by a lipid vesicle bilayer or when the particle diffuses at the lipid-monolayer interface of a droplet. For particles wrapped by membrane bilayers, we highlight the influence of the particle penetration depth and the impact of substructures on particle friction. Our work is complemented by hydrodynamic simulations that take into account the effects of the shape of the membrane wrapping the particle and the water gap separating the lipid bilayer membrane from the particle on translational particle drag. We show, however, that a purely hydrodynamic model is not suitable to describe the friction of a particle diffusing at the interface of an aqueous microdroplet in oil, stabilized by a single lipid layer. In hydrodynamic models, dissipation is solely described by the surface shear viscosity of the interface and the bulk fluid viscosity, but in this partial wetting configuration, an additional source of dissipation is required to account for fluctuations at the contact line. Hence, through experimental and numerical studies, we demonstrate that the dissipation contributions for the two geometries are fundamentally different.

\end{abstract}

\maketitle

\section{INTRODUCTION}

Interactions between biological cells and their surrounding medium, such as the internalization of foreign bodies through endocytosis, are crucial for the survival and development of living organisms \cite{Doherty2009, Mellman1996}. However, they are also associated with emerging sanitary threats, such as the ingestion and inhalation of microplastics or nanomaterials, which compromise the integrity of cell membranes \cite{Yong2020}. Therefore, understanding the mechanisms of interactions between micron-sized particles and lipid membranes is of paramount importance both for fundamental science and public health, yet, few experiments have been reported in the literature, and most of them focus on systems under very specific conditions.

Ultimately, a vesicle or a cell can be described in terms of the energetic barriers associated with its lipid bilayer: tension $\sigma$ and bending rigidity $\kappa_b$ resist membrane deformation and must be overcome for particle wrapping \cite{AgudoCanalejo2019}. When the adhesive energy between a particle and a membrane surpasses these contributions, it is spontaneously engulfed, with a penetration depth depending on the relative magnitudes of these energies.

Many biological processes are mediated by specific interactions between a ligand and a receptor (e.g., avidin-biotin complexes or antibody-antigen interactions \cite{Israelachvili2011}) and can be investigated using synthetic model systems composed of micron-sized particles and giant unilamellar vesicles (GUVs). For instance, van der Wel et al. \cite{vanderWel2017} explored membrane-mediated particle assemblies using avidin-coated particles and vesicles containing biotinylated lipids. Additionally, non-specific interactions, also prevalent in biological environments, have been examined through similar model systems. Spanke et al. studied the dynamics of particle wrapping by floppy vesicles as a function of PEG concentration, a depletion agent used to trigger adhesion (with adhesion energies of a few $\mathrm{\mu J}.\mathrm{m}^{-2}$) \cite{Spanke2020}. Their study is set in the context of the bending-dominated regime, where, $a < \lambda_\sigma$, ($a$ being the particle radius), with $\lambda_\sigma = \sqrt{\kappa_b/\sigma}$ as the bendocapillary length, capturing the balance between tension and bending rigidity. Ewins et al. demonstrated the electrostatic attraction between negative polystyrene particles and positively charged vesicles mainly composed of DOPC and a few percent of DOTAP, a cationic lipid, and showed that the penetration depth of the particle through the membrane increased with the amount of DOTAP \cite{Ewins2022}. Other studies have focused on investigating the interactions of polyelectrolyte-coated colloids with both small and giant lipid vesicles, as reported by Fery et al. \cite{Fery2003}.

Many theoretical and few experimental studies have been conducted to investigate the behavior of colloids wrapped by lipid membranes. In their pioneering work, Dietrich et al. \cite{Dietrich1997} demonstrated the impact of membrane tension on the penetration depth of latex particles, describing the equilibrium state with a partial wetting model based on tension and adhesion energies, but lacked precise control over the adhesion strength and did not quantitatively explore particle diffusion within the membrane. Dimova et al. \cite{Dimova1999} performed viscosimetry measurements by analyzing the motion of micron-sized particles on GUVs and derived membrane viscosity using a model assuming a spherical particle straddling a flat membrane at its equator, without accounting for local deformations or the presence of a water gap between the particle and the bilayer. Shigyou et al. \cite{Shigyou2016} studied the diffusion of submicron particles partially (wrapping degree unquantified) or fully engulfed by giant vesicles, where wrapping was obtained by applying strong centrifugation forces.
Theoretical and numerical studies \cite{Bahrami2014} have also shown that nanoparticle wrapping is strongly influenced by the particle-membrane adhesion potential range, particle shape, and global vesicle shape changes during wrapping. Theories \cite{AgudoCanalejo2015} have also been developed to describe the engulfment of nanoparticles by asymmetric membranes characterized by non-zero spontaneous curvature. Phenomena such as adhesion-induced segregation of membrane components have been considered, and engulfment diagrams have been established.

Numerous models have been developed to describe the motion of a membrane-bound particle. Saffman and Delbrück provided an analytical description of protein diffusion, considering a non-protruding cylindrical inclusion moving under the influence of Brownian motion in an incompressible fluid membrane, highly viscous compared to the surrounding bulk fluid \cite{Saffman1975}. Hughes, Pailthorpe and White refined this analysis by deriving the exact solution to the equations of motion for any combination of membrane and bulk fluid viscosities \cite{Hughes1981}. Later Evans and Sackmann extended this model by considering the presence of a rigid substrate in proximity to the membrane and revealed its influence on the transport of momentum into the third dimension \cite{Evans1988}. These models apply to cylindrical particles whose height equals the thickness of the membrane. When the particle is spherical and protruding in the bulk fluids, various approaches have been developed. Danov et al. numerically investigated the case of a spherical particle immersed in an interface where both shear and dilatation/compression effects are taken into account \cite{Danov1995} \cite{Danov2000}. Fischer et al., on the other hand, considered significant Marangoni effects across the interface, inducing its incompressibility \cite{Fischer2006}.

None of these models account for either the geometry of the membrane wrapping the particle or the water gap separating them. Nevertheless, they have often been used to interpret experiments involving particles partially wrapped by lipid membranes. In this paper, we characterize the translational friction experienced by the particle as a function of its penetration depth and compare our experimental data to a theoretical model built for a particle straddling a monolayer or a membrane bilayer. To obtain a better description of our system, we performed CFD numerical simulations considering three additional parameters: (i) the membrane segment wrapping the particle for three penetration depths, (ii) the presence of a water gap separating the particle and the lipid bilayer, and (iii) a solid substrate in their vicinity. We highlight that an increase of the water gap  significantly decreases the particle drag for relatively high membrane viscosities, but has a rather small effect for low values of the membrane viscosity. Our work also reveals that the presence of membrane substructures (e.g. tubes or buds) close to a particle contributes to increased drag. To discuss the origin and the quantitative measurement of the dissipation due to the lipid bilayers, we also measure the particle translational friction in the presence of a single monolayer. For the latter experiments, we encapsulate silica particles within water droplets in an oil (showing a viscosity similar to water) stabilized by a monolayer of lipids of the same nature. In the monolayer system, surprisingly we observe a significant increase of the particle friction, which is higher than that for particles wrapped by bilayers. This difference in particle drag can be explained considering the different geometries for monolayers and bilayers. In monolayers as in bare fluid interfaces, the particle is partially wetted by two fluids and the monolayer is not following the profile of the protruding particle. Hence, our experiments for monolayers can be explained by the partial wetting of the particle at the interface, where thermally excited fluctuations of the contact line add an extra friction contribution. Studying dissipative phenomena at the particle-membrane interface requires initiating adhesion. While particle wrapping by vesicles can occur spontaneously through specific or non-specific interactions, it can also be forced, e.g., by centrifugation. Inspired by the work of Ewins et al. \cite{Ewins2022}, we focus on the charge-induced interaction between micron-sized spherical particles with different zeta potentials and (positively charged) DOPC vesicles doped with 5 mol \% DOTAP, exploring regimes from no attachment to strong adhesion. Hence, to tune the adhesion we rely solely on the particle surface potential and the membrane tension, therefore avoiding the use of PEG depletants that are known to affect the membrane properties \cite{Lipowsky2013}. Specifically, we focus on negative silica particles and show that in this strong adhesion regime, an optically trapped particle brought close to a GUV experiences a force pulling it out of the trap and leading to its spontaneous engulfment. Theoretical studies and simulations \cite{Bahrami2014} have shown that in the bending-dominated regime, and where both spontaneous curvature and particle-membrane interaction potential range are equal to zero, only unbound and fully wrapped states are stable. Our study, considering high vesicle tension, nonzero potential range, and possible spontaneous curvature, explains the existence of intermediate wrapping states. Hence, this system allows a quantitative analysis of particle friction across a wide range of equilibrium states, from particle adhesion without optically resolvable membrane deformation, to full engulfment.

\begin{figure}
\centering
\includegraphics[width=0.9\linewidth]{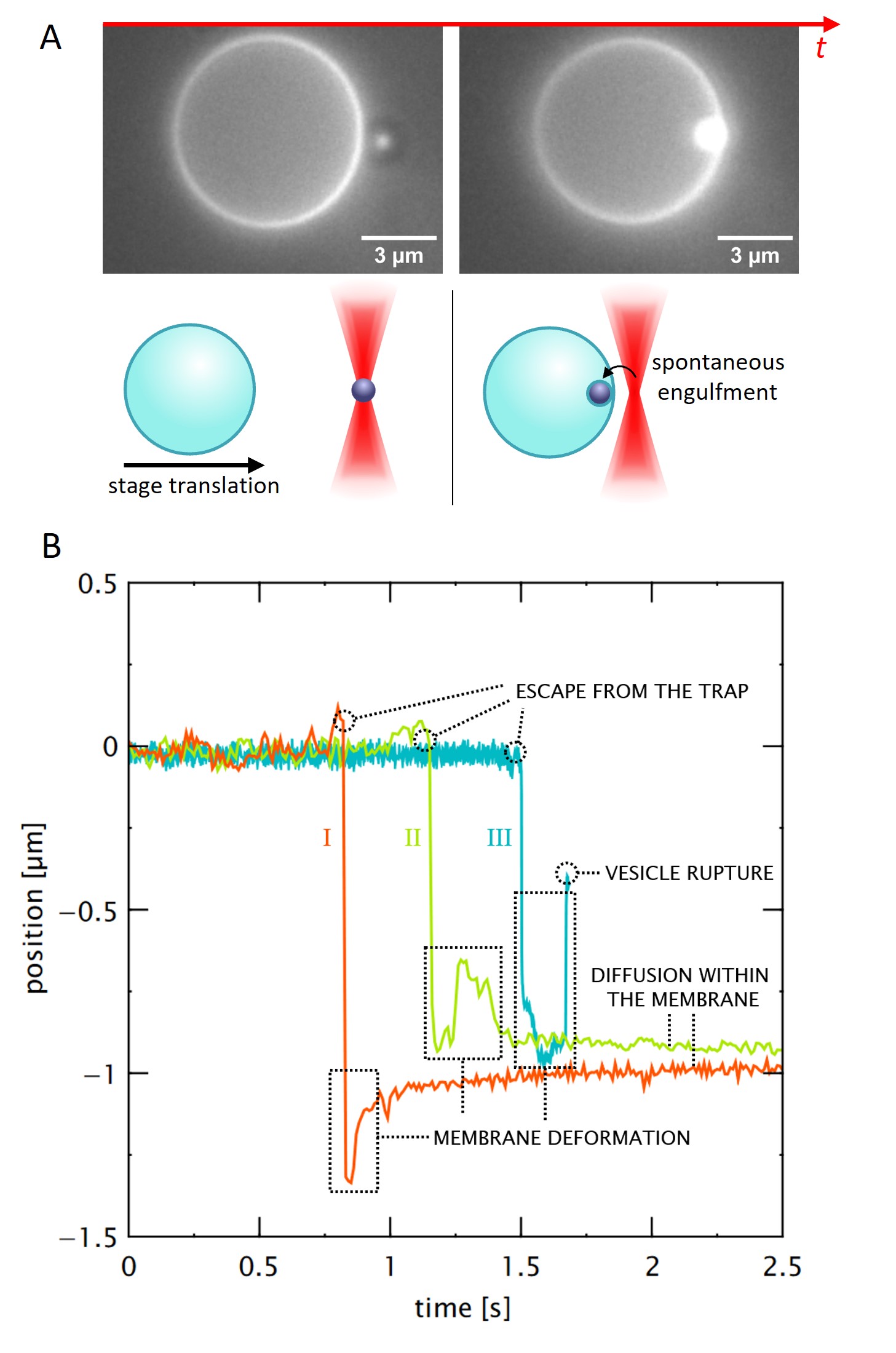}
\caption{(A) Combined fluorescence and bright-field microscopy snapshots of a silica particle (radius $a=0.6 \,\mathrm{\mu m}$)  initially trapped, escaping from the optical potential well of stiffness $\kappa_{x}=4.6 \,\mathrm{pN.\mu m^{-1}}$ and spontaneously penetrating inside a GUV (5 mol\% DOTAP). (B) Typical 1D trajectories of particles spontaneously escaping from the optical trap and penetrating inside a vesicle. The particles are initially confined within a potential well and subsequently extracted as they approach the membrane vicinity (distance below the optical resolution limit). During the wrapping process, the particles deform the membrane and reach an equilibrium position, with a fixed contact line (except for the curve III).}
\label{ParticleEscapeTrap}
\end{figure}

\begin{figure}
\centering
\includegraphics[width=1\linewidth]{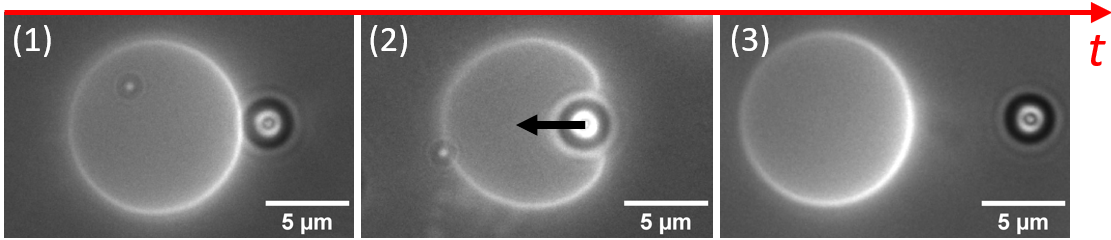} 
\caption{Combined fluorescence and bright-field microscopy snapshots showing a positively charged melamine formaldehyde particle (radius $a=1.25 \,\mathrm{\mu m}$) under optical trapping. The particle is brought into contact with the membrane but does not spontaneously escape the optical trap (1). A force (black arrow) is applied by translating the vesicle to induce engulfment, but the particle is not stably wrapped, and the deformation is temporary (2). When the vesicle is moved away from the particle using optical tweezers, no adhesion or membrane tube keeps the particle attached to the vesicle (3).}
\label{MF_particle_trap}

\end{figure}

\section{RESULTS AND DISCUSSION}

\subsection{Dynamics of particle engulfment by a lipid bilayer}

\begin{figure*}
\centering
\includegraphics[width=1\textwidth]{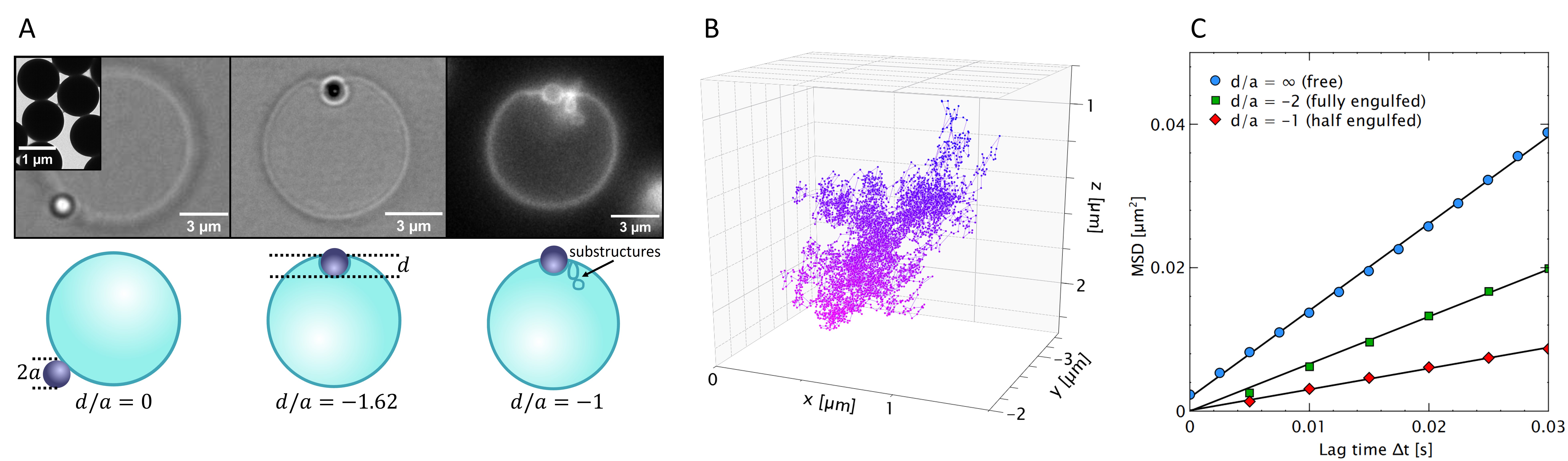} 
\caption{(A) Bright-field ($d/a = 0$ and $d/a = -1.62$) and fluorescence ($d/a = -1$) images of particle-vesicle systems showing varying penetration depths, accompanied by corresponding sketches. The vesicles contain 5 mol\% DOTAP and the particles  are bare silica microspheres. The inset is a transmission electron microscope (TEM) image of the silica particles. In the case of $d/a = -1$, internal membrane substructures are observed in the vicinity of the particle. (B) 3D reconstruction of the trajectory of a particle partially engulfed by a vesicle membrane and moving along its surface, from which the mean squared displacement (MSD) is calculated. (C) MSD as a function of lag time for free, fully and half engulfed silica particles. Translational diffusion coefficients $D_{tr}$  were extracted by fitting the data to the relationship $\mathrm{MSD} = 4D_{tr}\Delta t$ in the short lag time limit ($\Delta t < 0.03$ s).}
\label{DegreeEngulfmentMSD}
\end{figure*}

In order to characterize the wrapping dynamics of the membrane on the particle, bare silica microspheres (radius $a=0.6 \,\mathrm{\mu m}$) were mixed with DOPC GUVs containing 5 mol\% DOTAP. An optical trapping system was then used to bring a particle into the vicinity of a membrane. Most of the time, beyond a critical distance, the attractive force exerted by the vesicle exceeds the restoring force $F_{x} = \kappa_{x}x$ of the trap. This causes the particle to escape and subsequently enter the vesicle, reaching an equilibrium position with a fixed contact line. A similar behavior was also observed by Dietrich et al. \cite{Dietrich1997}. Fig. \ref{ParticleEscapeTrap}A shows an example of a particle initially optically trapped in a harmonic potential of stiffness $\kappa_{x}=4.6 \,\mathrm{pN.\mu m^{-1}}$, spontaneously and completely engulfed by a vesicle. 

The wrapping dynamics strongly depends on the membrane tension, as illustrated in Fig. \ref{ParticleEscapeTrap}B, which displays typical trajectories of particles escaping from the optical trap and penetrating inside a GUV. The curve I shows a particle inducing a localized deformation associated with the opening of a neck, without visible global deformation. About 60 ms after intake, the neck formed by the wrapping of the particle stabilizes (rebound \cite{Spanke2022}), and the particle reaches an equilibrium position where the particle-membrane contact line is pinned. The particle is completely engulfed by the vesicle, as evidenced by the distance before/after engulfment, approximately equal to $2a$. The curve II indicates a more complex global-scale deformation induced by partial engulfment of the particle, followed by stabilization of the contact line. Finally, the curve III illustrates a globally visible optical deformation of the GUV induced by the particle. About 170 ms after intake, the vesicle relaxes back to its original spherical shape. Partial wrapping of the particle by the membrane leads to an increase in vesicle tension up to its lysis tension, causing its rupture (likely through pore formation). Thus, the particle does not reach an equilibrium position within the membrane in this configuration. 

Although we could not quantitatively measure the membrane tension for each vesicle, qualitative analysis of vesicle deformation during particle engulfment reveals that significant, globally uniform deformations (i.e. without micrometric propagation of membrane distortion), indicative of high membrane tension ($\sigma > 10^{-6} \,\mathrm{N.m^{-1}}$), lead to a shallower penetration depth compared to vesicles experiencing only local deformations optically undetectable (only related to the opening of a neck). In low-tension vesicles ($\sigma < 10^{-6} \,\mathrm{N.m^{-1}}$), subject to thermal shape fluctuations detectable by optical microscopy, the opening of a neck does not induce global deformation due to the area reservoir contained within these undulations. 

For tensionless vesicles, resistance to the strong adhesion between the particle and the membrane arises solely from bending constraints, usually leading to complete particle engulfment, however for high-tension GUVs, the particle may undergo partial engulfment. In this case, the particle-vesicle configuration is determined by minimizing the total energy of the system $E = E_w + E_\sigma$, with $ E_w = 2\pi adw$ and $ E_\sigma = \pi d^{2}\sigma$ representing the adhesion and tension energies, respectively \cite{Deserno2004}. Thus the penetration depth $d/a$ (defined in Fig. \ref{DegreeEngulfmentMSD}A) depends solely on the balance between the particle-membrane adhesion energy $w$, and the tension of the vesicle $\sigma$ with $w = -(d/a)\sigma$.

To monitor particle dynamics, measurements at 900 FPS were conducted. However, even at such high acquisition frequency, precise particle tracking is not feasible, and the rate at which it escapes the trap only constitutes a lower limit, with a velocity $U > 440\, \mathrm{\mu m.s^{-1}}$ (curve III) and an associated Stokes' drag force $F_{drag} = 6\pi\eta_{w} a U > 5\,\mathrm{pN}$ (with $\eta_{w} = 10^{-3} \ \mathrm{Pa.s}$ the viscosity of water). 

As mentioned earlier, a low-tension vesicle, highly fluctuating, has enough reservoir area to accommodate the uptake of one or more particles \cite{Fessler2023}. When a vesicle assumes a perfectly spherical shape, with no thermally activated membrane undulations, any further increase in its volume is significantly constrained by its stretching elasticity, limiting bilayer expansion to a few percent before rupture \cite{Lipowsky2019}. Thus, the engulfment of a particle into a vesicle increases membrane tension (especially in small vesicles), potentially leading to its disruption. A vesicle rupturing after the uptake of a particle indicates that its tension was initially very high, close to its lysis tension. The condition for half engulfment of a particle by the membrane is $w  \ge \sigma$, thus, considering a typical lysis tension value of $\sigma \approx 10^{-2}\,\mathrm{N.m^{-1}}$ \cite{Olbrich2000},  the adhesion force $F_{w} \sim aw$ can reach magnitudes on the order of several nN, with $w = 10 \, \mathrm{mJ}.\mathrm{m}^{-2}$, similar to the surface energy of the biotin-avidin interaction \cite{Moy1999}. This analysis highlights the strong adhesion at play, resulting in the wrapping of the particle even in the high tension regime. This contrasts with other studies, such as \cite{Fery2003}, where the wrapping of polyelectrolyte-coated beads by zwitterionic vesicles was observed only for highly fluctuating membranes.

The wrapping trajectories reported by Spanke et al. \cite{Spanke2022} for particles engulfed by tensionless vesicles (in the bending-dominated regime), by depletion effect, are similar to those observed in our low-tension vesicle experiments (curve I), where the particle rebounds before settling at its final depth. However, the wrapping velocities differ significantly: a few $\mathrm{\mu m.s^{-1}}$ in their case compared to at least several hundred $\mathrm{\mu m.s^{-1}}$ in ours. This suggests that the interaction mechanism (depletion vs. electrostatics) and the associated energies do not substantially influence the overall shape of the wrapping trajectories. In contrast, curves II and III exhibit qualitatively distinct behaviors due to high adhesion and vesicle tension.

To assess whether particle attraction is primarily driven by surface charge, we tested different chemical compositions: silica, polystyrene, aminated silica and melamine formaldehyde. For each material, the particles were brought close to vesicles using optical tweezers, and we observed whether the particle was spontaneously pulled out of the trap by the GUV. Highly negative silica and polystyrene particles were almost systematically engulfed spontaneously (often leading to full wrapping), whereas positive melamine resin particles showed no adhesion, even when a force was applied using the trap (see Fig. \ref{MF_particle_trap}). Aminated silica particles, weakly negatively charged, exhibited intermediate behavior, with spontaneous engulfment occurring less frequently (see Table \ref{tab:engulfment}). Adhesion is thus mainly governed by electrostatics and can be controlled by varying particle surface potential. 

\begin{table}[h!] 
\caption{\label{tab:engulfment}Correlation between spontaneous engulfment by positively charged vesicles ($\zeta_{\mathrm{DOTAP}} = +33$ mV) and particle zeta potential.}
\footnotesize 
\begin{ruledtabular}
\begin{tabular}{lcc}
\textrm{Particle material} &
\textrm{$\zeta$ (mV)} &
\textrm{Spontaneous engulfment\footnote{\textquotedblleft Yes\textquotedblright\ indicates $> 90\%$ of particles were spontaneously engulfed, \textquotedblleft No\textquotedblright\ indicates none, and \textquotedblleft Occasional\textquotedblright\ indicates $< 40\%$.}} \\
\colrule
Silica & -63 & Yes \\
Polystyrene & -54 & Yes \\
Aminated silica & -12 & Occasional \\
Melamine formaldehyde & +25 & No \\
\end{tabular}
\end{ruledtabular}
\end{table}

\subsection{Influence of penetration depth on particle friction in a lipid bilayer}

\subsubsection{Experiments}

In this section, we aim to describe the equilibrium dynamics of particle-vesicle systems. A distribution of tension and adhesion energy allowed us to observe various penetration depths, described by a dimensionless parameter $d/a$, where $a$ corresponds to the particle radius and $d$ the distance between the immersed pole of the particle and the membrane surface, as shown in Fig. \ref{DegreeEngulfmentMSD}A. 

Particles exhibit a pinning of the contact line and show anisotropic motion constrained by the curvature of the GUV. Therefore, it is necessary to project the experimentally observed 2D path into a 3D trajectory to accurately describe the diffusion of the particle on the vesicle surface. This trajectory reconstruction assumes that the vesicle has a spherical shape, which is an approximation given that gravitational effects and adhesion to the substrate may cause a flattening of the vesicle. Nevertheless, this approximation is quite suitable as long as the particle does not stray too far from the focal plane of observation where the radius of curvature is measured, which is the case for our system. Fig. \ref{DegreeEngulfmentMSD}B represents the 3D reconstruction of the diffusion of a partially engulfed particle, based on its apparent 2D trajectory. The mean squared displacement was calculated using the method of Shigyou et al. (see section \ref{subsec:AcquisitionTracking}) and examples of MSD corresponding to free, fully and half engulfed particles are plotted in Fig. \ref{DegreeEngulfmentMSD}C.

\begin{figure}
\centering
\includegraphics[width=1\linewidth]{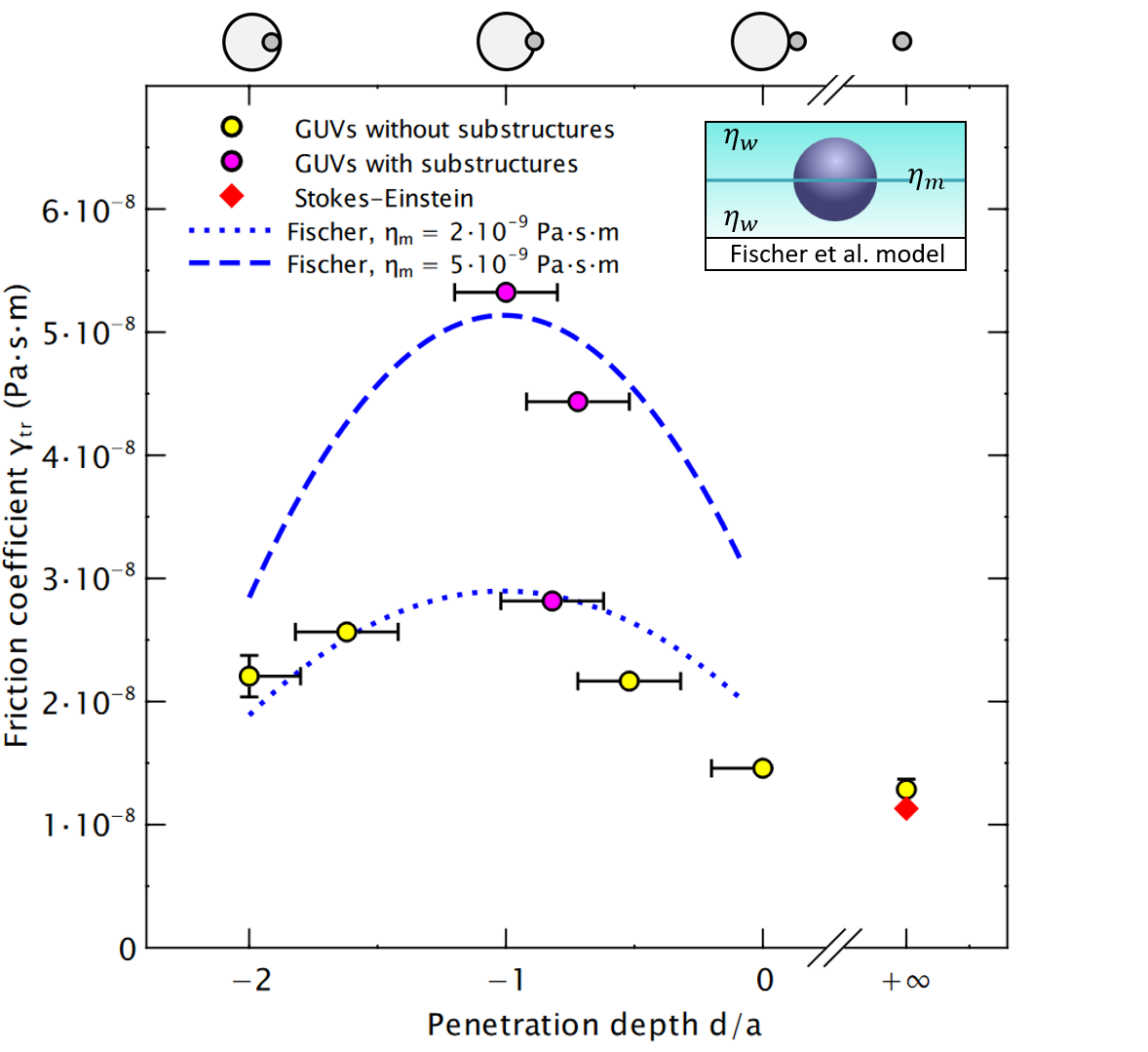} 
\caption{Translational friction coefficient $\gamma_{tr} = k_{B}T/D_{tr}$ of a silica particle versus penetration depth $d/a$ inside a 5 mol\% DOTAP GUV membrane (radius ranging from 3.5 to 6 µm). Number of particles analyzed: $n = 14$. The yellow circles correspond to experimental data for particle-vesicle systems without substructures, while the pink circles represent systems with substructures. The dashed and dotted blue curves correspond to a hydrodynamic model \cite{Fischer2006} for two different membrane viscosities. The red diamond corresponds to the theoretical value of particle diffusion calculated using the Stokes-Einstein relation. Inset: sketch of the hydrodynamic model \cite{Fischer2006}.}
\label{FrictionVsPenetrationDepth}

\end{figure}

In addition to the various degrees of engulfment observed experimentally, some vesicles may also contain substructures such as lipid aggregates, filaments, smaller vesicles, or may show some multilamellar regions, which could increase the friction experienced by the particle. Fig. \ref{FrictionVsPenetrationDepth} summarizes these effects and depicts the evolution of translational friction between a silica particle and a GUV containing 5 mol\% DOTAP. Yellow circles correspond to particles engulfed by vesicles without substructures observable by fluorescence microscopy, while pink circles correspond to cases where substructures are present (see Fig. \ref{DegreeEngulfmentMSD}A for $d/a = -1$, the vesicle exhibits substructures near the particle). The dashed and dotted blue curves are associated to numerical calculations by Fischer et al. describing the friction of a sphere passing through an incompressible two-dimensional liquid of viscosity $\eta_{m}$ (membrane), separating two liquids of infinite depth and viscosity $\eta_{w}$ (water) \cite{Fischer2006}. For $-2 < d/a < 0$ the friction force associated with the translational motion of the sphere at velocity $U$ is defined by:
\begin{equation}
    F_{drag} = k_{tr}\eta_waU
\end{equation}
with the translational drag coefficient $k_{tr} = 6 \pi$ when the sphere is far away from the interface $(d\to +\infty)$. To account for the interfacial effects, $k_{tr}$ is expressed in linear order of the Boussinesq number: $k_{tr} = k^{(0)}_{tr} + \mathscr{B}k^{(1)}_{tr}  + o(\mathscr{B}^2)$, with $\mathscr{B} = \eta_{m}/(2\eta_{w}a)$. In this configuration, the coefficient $ k^{(0)}_{tr}$ is defined as follows:

\begin{equation}
k^{(0)}_{tr} = 6\pi + \frac{3\pi /2}{1+(d/a+1)^{2}}
\end{equation}

and where the first-order correction term $k^{(1)}_{tr}$ does not have a defined analytical expression.

\begin{figure*}
\centering
\includegraphics[width=1\textwidth]{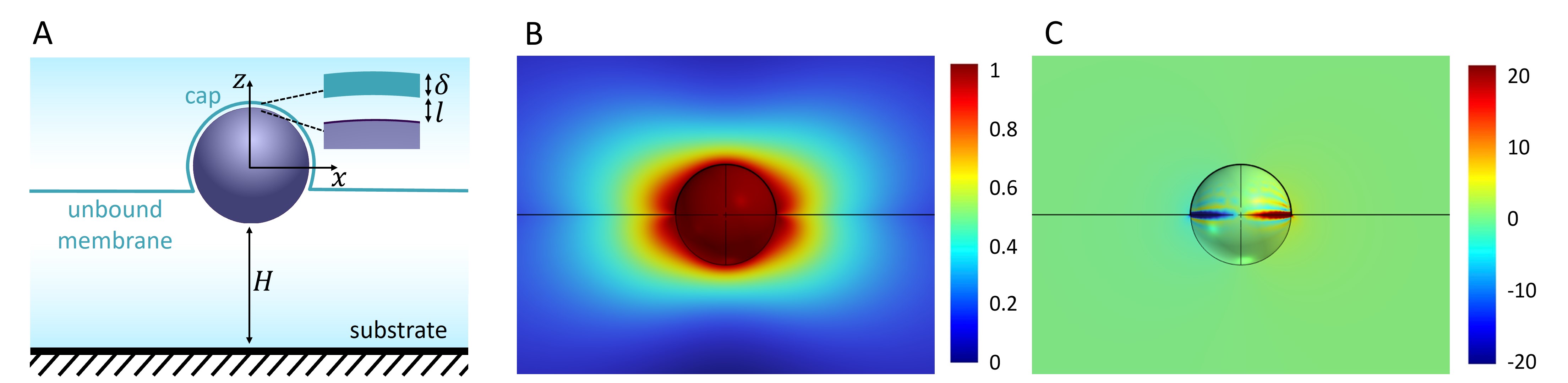}
\caption{(A) Sketch of the simulation domain geometry. The membrane, defined by a constant surface shear viscosity $\eta_{m}$, has a thickness $\delta = 5\,\mathrm{nm}$ and is separated from the particle surface by a water gap of thickness $l$. A rigid flat substrate is placed opposite to the membrane at a distance $H$ from the particle surface. (B) Dimensionless velocity and (C) pressure fields for $d/a = -1$, $l = 5\ \mathrm{nm}$ and $\eta_{m}=2\cdot 10^{-9}\ \mathrm{Pa.s.m}$. The velocity is made dimensionless by $U$ and the pressure by $ \eta_{w} U/a$.}
\label{GeometrySimulations}
\end{figure*}

This model assumes no rotation for a particle in contact with the membrane and no deformation of the interface, remaining flat for every degree of the particle penetration (see inset Fig. \ref{FrictionVsPenetrationDepth}). Dissipation results from the bulk viscosity seen by the spherical particle showing a size significantly larger than the membrane thickness and by the viscosity of the flat membrane, which forms a hole to accommodate the particle inclusion. Therefore, the cases $d/a = -2$ and $d/a = 0$ are identical and the friction follows a parabolic form. However, in an endocytosis-like process, a particle does not perforate the bilayer but induces deformation, forcing the membrane to locally match the particle shape. This symmetry breaking is not considered in current models and may explain the deviations observed between our experimental data and the numerical calculations. Note that although a discrepancy exists, experimental results can be reasonably well fitted by the hydrodynamic model \cite{Fischer2006} with a viscosity $\eta_{m}=2\cdot 10^{-9}\ \mathrm{Pa.s.m}$, both showing a maximum friction when the spherical particle is symmetrically immersed, with its equator coinciding with the membrane plane. However, some data cannot be fitted with this value of $\eta_{m}$ and correspond to particles interacting with vesicles containing substructures, which may increase the viscosity of the membrane. 

For fluid bilayers, the molecular rearrangement of lipids prevents the generation of shear stress across the membrane, thus the latter can be described as a thin region of liquid with viscosity $\eta$ and thickness $\delta$, without generation of velocity gradient, or equivalently, can be treated as an anisotropic material with a constant surface shear viscosity $\eta_m = \delta \eta$. Therefore, the increase in friction between particles and GUVs containing substructures in their vicinity can be attributed to the linear dependence of surface shear viscosity on membrane thickness, and hence on the number of lipid monolayers. Consequently, the two highest values of the friction (pink circles), deviating from the trend ($\eta_{m}=2\cdot10^{-9}\ \mathrm{Pa.s.m}$), can be described by a (2.5 times) higher effective viscosity, possibly indicating the presence of a second bilayer. 

Relating our findings to previous studies, Shigyou et al. \cite{Shigyou2016} studied the diffusion of submicron polystyrene particles wrapped by DOPC GUVs (by centrifugation) and classified the wrapping states into two categories: "full" or "partial". Due to the particle size (200 - 800 nm in diameter), intermediate wrapping states could not be optically resolved. They reported a higher drag for fully wrapped particles compared to partially wrapped ones, hence, by comparing with our data (although involving larger particles), this suggests that the partially wrapped particles in their study likely had very shallow penetration depths. In order to estimate membrane viscosity, their data were compared with the DADL model \cite{Danov1995} \cite{Dimova1999}, which assumes a flat membrane crossing the particle equator and does not account for local deformations around the particle or the presence of a water gap. Although this model is a simplified approach, they determined a membrane viscosity of $\eta_{m}=2\cdot10^{-9}\ \mathrm{Pa.s.m}$, the same value we found in our experiments. Recently, Van der Wel et al. \cite{vanderWel2017} studied the adhesion between avidin-coated particles and vesicles containing biotinylated lipids, and observed two equilibrium states: fully wrapped or attached. Surprisingly, they reported higher diffusivity for fully wrapped particles, and a possible explanation for this result was qualitatively attributed to an increased membrane-particle distance compared to the attached state. However, these findings contrast with our results, where fully wrapped particles ($d/a = -2$) exhibit slower diffusion (higher friction) compared to attached particles ($d/a = 0$) (see Fig. \ref{FrictionVsPenetrationDepth}). In their study, the DADL model \cite{Danov1995} was also applied to extract the membrane viscosity, yielding a value of $\eta_{m}=1.2\cdot10^{-9}\ \mathrm{Pa.s.m}$. The impact of the particle-membrane distance, discussed in \cite{vanderWel2017}, which may be described as a water gap, is quantitatively analyzed in the following section.

\subsubsection{Simulations}

\begin{figure}
\centering
\includegraphics[width=1\linewidth]{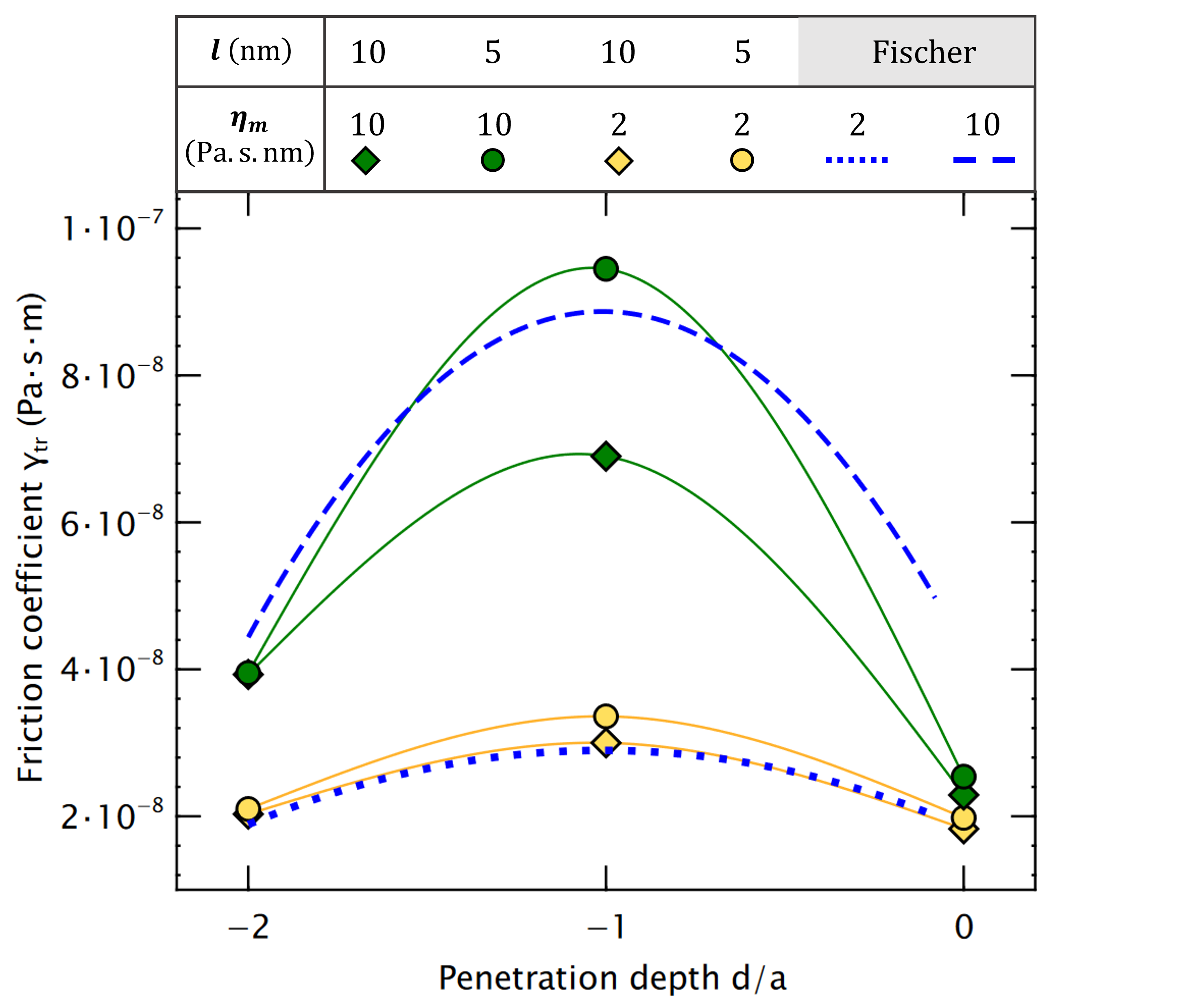} 
\caption{Translational friction of a spherical particle (radius $a=0.6 \,\mathrm{\mu m}$) as a function of penetration depth through a membrane, modeled as a layer of incompressible fluid (width $\delta = 5\,\mathrm{nm}$) and constant surface shear viscosity $\eta_{m}$, separated by two identical regions of incompressible liquid (water). A water gap (thickness $l$) between the particle and the membrane is included. Our numerical results are compared to Fischer et al. calculations for surface shear viscosites $\eta_{m}=2\cdot 10^{-9}\ \mathrm{Pa.s.m}$ and $\eta_{m}= 10^{-8}\ \mathrm{Pa.s.m}$ (blue dotted and dashed curves). The points from the simulation, represented as circles and diamonds, are connected by a solid curve serving as a visual guide and does not represent an interpolation.}
\label{Simulations_VS_Ficher}

\end{figure}

To account for the hydrodynamic interactions between the colloidal particle and the membrane we ran CFD numerical simulations considering a free-buoyant rigid sphere with radius $a=0.6 \,\mathrm{\mu m}$ suspended in an incompressible Newtonian liquid at different distances from a vesicle. The geometry of the simulation domain is sketched in Fig. \ref{GeometrySimulations}A. The membrane wraps the particle and is described as a non-deformable (the shape remains unchanged throughout the simulation) and incompressible fluid layer with width $\delta = 5\,\mathrm{nm}$ and constant surface shear viscosity $\eta_{m}$. A Cartesian reference frame with the origin at the sphere center and with axes as indicated in the figure is considered. The particle-membrane distance is constant and denoted by $l$. A rigid flat wall is placed opposite to the membrane at a distance $H$ from the particle surface. This distance is set to $H = 2.1 \,\mathrm{\mu m}$ corresponding to the average value estimated from the experiments (using Faxén's hydrodynamic calculations for the translational drag of a spherical particle as a function of its distance from a solid wall \cite{Faxn1922}). Fig. \ref{GeometrySimulations}B and C, for instance, represent the dimensionless velocity and pressure fields for $d/a = -1$, $l = 5\ \mathrm{nm}$ and $\eta_{m}=2\cdot 10^{-9}\ \mathrm{Pa.s.m}$. The two limiting cases are $d/a = 0$ where the membrane is completely flat and $d/a = -2$ where the particle is fully embedded by the membrane. Flow continuity is imposed at the membrane interfaces and no-slip condition at the external domain boundaries (including the physical wall). To evaluate the translational drag coefficient $\gamma_{tr}$, a constant velocity $U$ along the $x$-direction is applied at the particle surface and the force is calculated. Due to symmetry, only one-half of the domain is considered (the $xz$-plane is the symmetry plane). Assuming negligible inertial forces, the fluid dynamics equations in the two domains, i.e., water ($w$) and membrane ($m$), read: 

\begin{equation}
\left\{
    \begin{aligned}
        \nabla \cdot \boldsymbol{v_i} &= 0 \\
        \nabla \cdot \boldsymbol{\sigma_i} &= -\nabla p_{i} + \eta_{i}\nabla^{2}\boldsymbol{v_i} = \textbf{0}
    \end{aligned}
\right.
\quad i = w, m
\end{equation}

where $\boldsymbol{\sigma_i} = -p_{i}\boldsymbol{I} + 2\eta_{i}\boldsymbol{D_{i}}$ is the stress tensor on the i-th fluid, with $\boldsymbol{D_{i}}$ the rate-of-deformation tensor, $\boldsymbol{I}$ the unity tensor, and $v_{i}$, $p_{i}$ and $\eta_{i}$ are the velocity, pressure, and viscosity of the i-th fluid, respectively. The system of equations is solved using the finite element method within a sufficiency large box to ensure a negligible boundary effect. Numerically, we found that a box (L) greater than 20 times the radius of the sphere suffices to verify the above statement. The mesh, created with tetrahedral elements, is non-uniform, to allow an accurate description of the different characteristic lengths of the problem. Specifically, boundary layers are added between the particle and the membrane due to the very small gap. Mesh convergence was verified for all calculations presented in this work. Depending on the geometry, the total number of elements varies between approximately 130000 and 150000.

Several studies have demonstrated the existence of a variable water gap thickness in solid surface-membrane systems. For instance, Rädler et al. \cite{Rdler1995} investigated the weak interaction of giant vesicles with flat rigid substrates and found a mean separation distance of about 40 nm. Other studies have reported the existence of a 1-3 nm water gap in supported bilayer systems \cite{Johnson1991} \cite{Bayerl1990} (given the strong adhesion in our system, the particle-membrane equilibrium distance is likely near the lower bound of this range). Concerning viscosity, several authors have measured values for DOPC bilayers ranging from 1.2 to 15 $\cdot 10^{-9}\ \mathrm{Pa.s.m}$ \cite{vanderWel2017} \cite{Shigyou2016} \cite{Faizi2022} \cite{Hormel2014}. Fig. \ref{Simulations_VS_Ficher} shows the calculated values of the translational friction coefficient as a function of penetration depth for different water gap thicknesses and membrane viscosities, reflecting the range of data obtained in these different studies, with a water gap of 5 – 10 nm and viscosity ranging from 2 to 10 $\cdot 10^{-9}\ \mathrm{Pa.s.m}$.

Our simulations exhibit a trend similar to Fischer et al. calculations, with maximum friction at half-engulfment. However here, the deformed shape of the membrane induced by particle wrapping is taken into account and the evolution of friction is no longer symmetrical with respect to the position $d/a = -1$. These simulations provide a correction to the hydrodynamic model \cite{Fischer2006}, allowing a more accurate description of our experimental results, also displaying asymmetry. In our simulations, the presence of a water gap separating the particle and the membrane was also taken into account. A reduction in the thickness of this water layer results in increased friction, more pronounced when the membrane is highly viscous. This variation is minor for a viscosity $\eta_{m}=2\cdot 10^{-9}\ \mathrm{Pa.s.m}$ but becomes significant for a viscosity $\eta_{m}=10^{-8}\ \mathrm{Pa.s.m}$. In our experiments, with a lipid bilayer viscosity estimated at $\eta_{m}=2\cdot 10^{-9}\ \mathrm{Pa.s.m}$, the thickness of the water gap has thus a small effect on friction. Hence, dissipation primarily occurs in the regions close to the junction between the unbound membrane plane and the cap (formed by the particle portion engulfed and covered by the membrane, see Fig. \ref{GeometrySimulations}A).

\begin{figure*}
\centering
\includegraphics[width=1\linewidth]{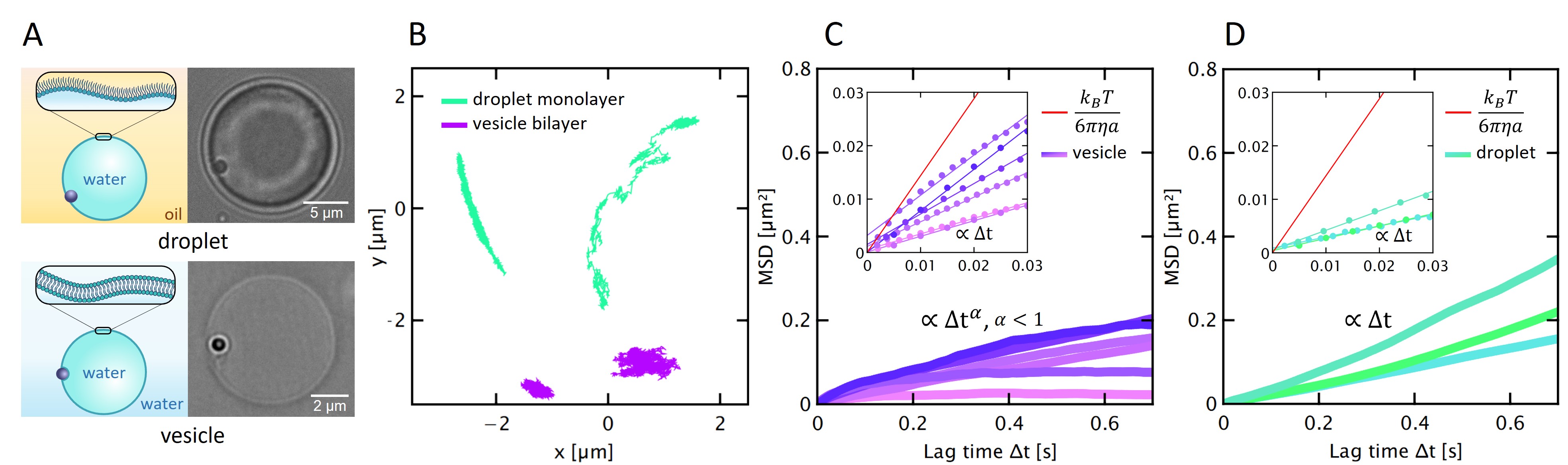} 
\caption{(A) Schematic representations and bright-field images of a particle wrapped by a vesicle bilayer and a particle at the interface between the internal aqueous phase and the external oil phase of a droplet, stabilized by a lipid monolayer. The lipid composition used for the fabrication of the vesicles and droplets is identical. (B) Typical two-dimensional projected (in the (x,y) plane) particles trajectories in the two configurations, for an acquisition time of 10 s and vesicle/droplet radii ranging from 3.5 to 7 µm. (C) Mean squared displacement (calculated from the 3D reconstruction of the 2D projected trajectory of the particle) as a function of lag time for particles wrapped by a lipid bilayer, with different penetration depths and (D) particles wetting a lipid monolayer. In the case of particles encapsulated in droplets, the MSD scales linearly with time, even for longer timescales ($\Delta t \gg 0.03$ s), characteristic of purely Brownian motion, but exhibits a lower diffusion coefficient compared to particles wrapped by a bilayer (see insets). In this latter case, the transport is hindered and manifests as a sublinear increase of the MSD at longer timescales ($\mathrm{MSD} \propto \Delta t^{\alpha}$, with $\alpha = 0.78 \pm 0.14$ based on all the data analyzed). The diffusion is systematically lower in both cases compared to the Stokes-Einstein predictions for a sphere in the bulk (red curve).}
\label{dropletVSGUV}
\end{figure*}

\section{Particle diffusion at the water/lipid monolayer/oil interface}

The analysis of the diffusion of particles wrapped by GUVs, potentially containing substructures near their contact zone, indicates an influence of the effective membrane thickness on translational friction. To investigate further the effect of the layer thickness, we performed experiments using lipid monolayers showing a thickness that is half that of lipid bilayers. We encapsulated particles in water droplets stabilized by a lipid monolayer (same composition as the vesicles) and dispersed in oil (dodecane, $\eta_{o} = 1.324 \cdot 10^{-3}\ \mathrm{Pa.s}$ \cite{Iwahashi1990}). 

Unlike GUVs, where particles spontaneously adhere to the outer leaflet of the lipid bilayer during the observation phase, here, the particles are directly encapsulated during the fabrication process. Fig. \ref{dropletVSGUV}A illustrates the two systems: at the bottom, a particle partially engulfed by a vesicle bilayer and at the top, a particle encapsulated in a droplet, in close proximity with the monolayer. 

In pure water, the particle does not adhere to the interface and diffuses freely in three dimensions within the aqueous droplet. The addition of NaCl in the aqueous phase (10 mM) results in a reduction of the Debye length and causes the particle to approach the interface. 
In this configuration, the particle does not deform the lipid monolayer but is in a state of partial wetting, moving along the curvature of the droplet and exploring a larger region than particles wrapped by a GUV bilayer. Fig. \ref{dropletVSGUV}B depicts the projected 2D trajectories of particles encapsulated in droplets and wetting the interface (green) and of particles partially engulfed by GUVs (purple), over the same acquisition time (10 s), for vesicle/droplet radii ranging from 3.5 to 7 µm, clearly showing this phenomenon. 

Although the trajectories of particles encapsulated in droplets are more extensive, the mean squared displacement evaluated at short times (MSD linear with the lag time for $\Delta t < 0.03$ s) reveals slower diffusion compared to particles wrapped by lipid bilayers. MSD analysis for particle-bilayer systems at longer times reveals a sublinear behaviour characteristic of confined diffusion dynamics (see Fig. \ref{dropletVSGUV}C and D). The crossing from the diffusive and subdiffusive regimes occurs around $\Delta t \simeq 0.05$ s, and the subdiffusion is characterized with an exponent $\alpha = 0.78 \pm 0.14$ (no correlation was observed between $d/a$ and $\alpha$). This phenomenon is ubiquitous in cellular biology and has also been studied using model systems. Several works have shown that the presence of substructures (buds) \cite{Shigyou2016} and membrane crowding induced by proteins \cite{Horton2010} or lipid domains \cite{Skaug2011} lead to anomalous transport. In our system, the vesicles are made of only two types of lipids in the liquid phase (and 1\% of a fluorescently labeled lipid), and thus diffusion cannot be perturbed by complex systems as proteins. The presence of substructures may explain this type of long time diffusion, but in our system, it appears even in GUVs without optically visible substructures. The finite-size effects induced by the spherical shape of the vesicle are not the cause of this confinement either, as it does not occur for a particle encapsulated in a droplet of the same size. Another hypothesis explaining particle subdiffusion could be the formation of an electrostatically induced DOTAP-rich zone in its vincinity. Among the various factors possibly contributing to particle subdiffusion, the role of membrane elasticity has also been investigated. Daddi-Moussa-Ider et al. analytically demonstrated that the motion of a particle near a membrane causes its deformation, followed by relaxation, generating a flow acting back on the particle with a certain delay. This memory effect, induced by the elastic nature of the membranes, is responsible for a transient subdiffusive phenomenon \cite{DaddiMoussaIder2016}. Although their work describes the diffusion of a particle near a membrane rather than being wrapped, this interpretation remains compelling in explaining the experimentally observed anomalous diffusion.

Going back to the short lag time diffusive regime, we measured a particle diffusion at the lipid monolayer interface significantly lower than for lipid bilayers even if the thickness of the monolayer is half that of the bilayer. Using the hydrodynamic model \cite{Fischer2006} to fit our experimental results for monolayers yields a surface shear viscosity of approximately $1.5\cdot 10^{-8}\ \mathrm{Pa.s.m}$, which is substantially higher than the estimated viscosity of lipid bilayers, even those containing substructures (in the range $\eta_{m}=2 \cdot 10^{-9}-5\cdot 10^{-9}\ \mathrm{Pa.s.m}$, see Fig. \ref{FrictionVsPenetrationDepth}).

In this model and our simulations, the interface is described as flat (no roughness), since interfacial tension stresses dominate over viscous stresses, with a capillary number $Ca = \eta U/\sigma_{wo} \ll 1$ where $U$ is the particle velocity and $\sigma_{wo}$ is the water-oil interfacial tension. The triple-phase contact line between the particle and the two immiscible liquids appears stationary at the microscopic scale, but at the molecular level, it undergoes significant thermal motion, leading to additional energy dissipation. To more precisely capture this behavior, we used the approach of Boniello et al. \cite{Boniello2015} where a spherical microparticle diffuses at the water/air interface with a contact angle $\theta$ (related to the penetration depth by $1+\cos\theta = -d/a$), satisfying the local equilibrium of the triple contact line. Thermally activated displacements of the interface at the contact line induce additional random forces acting on the particle, which are associated with an additional viscous friction through the fluctuation-dissipation theorem. The triple-phase contact line fluctuates around its average position with a characteristic deformation length scale denoted as $\lambda$, inducing a line friction acting on the particle surface $\gamma_{tr,L}$ defined by:

\begin{equation}
    \gamma_{tr,L} = \frac{\pi a \sin \theta}{2 k_{B}T \lambda}\left[ \sigma_{wo}\lambda(1- \cos \chi) \right]^{2}\tau
\end{equation}

\begin{figure}
\centering
\includegraphics[width=0.8\linewidth]{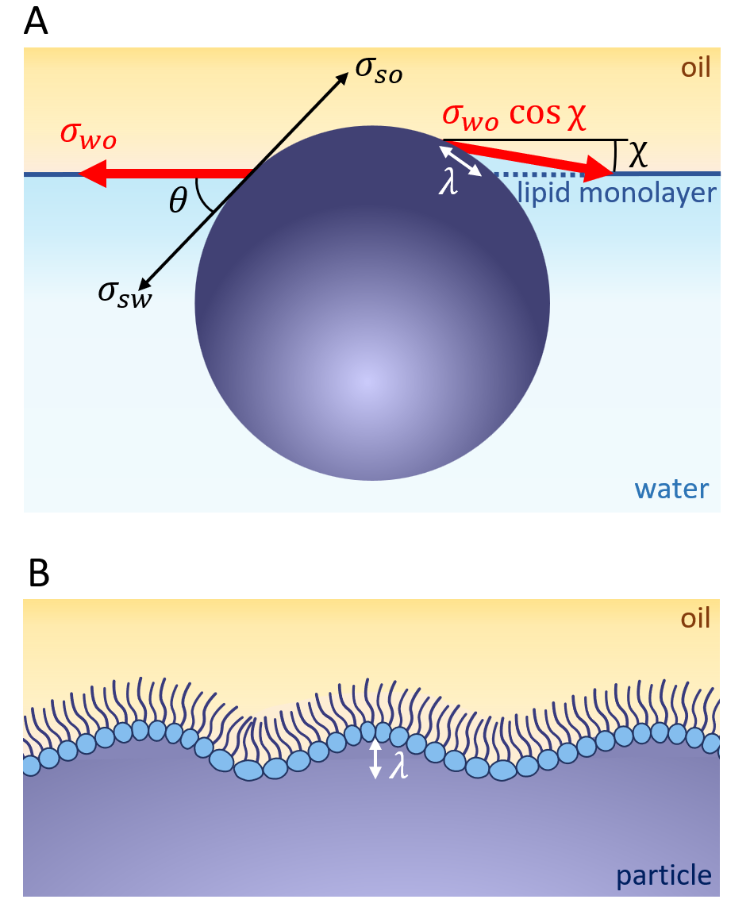} 
\caption{Sketch of a particle diffusing at the water/lipid
monolayer/oil interface. (A) $\theta$ corresponds to the angle between the horizontal water/oil interface and the tangents to the particle surface at the interface. $\lambda$ describes the typical fluctuation length of the interface, and $\chi$ is defined by the angle between the horizontal plane and the tangents of the interface on the particle surface. (B) Schematic representation of the solid/liquid/liquid triple line fluctuations.}
\label{DropletParticle}

\end{figure}

with $\chi = \arctan ({4 \sin \theta/\pi})$ being the angle between the horizontal plane and the tangents of the interface on the particle surface and $\tau = \kappa_0^{-1}$ corresponding to the correlation time of the interfacial fluctuations. This correlation time differs from that applied in the article by Boniello et al. describing a water/air interface. Here, both media are liquids with viscosities $\eta_{w}$ (water) and $\eta_{o}$ (oil) and the equilibrium frequency of random molecular displacements, $\kappa^0$, is defined by \cite{blake1993dynamic}:

\begin{equation}
    \kappa_0 = \frac{h k_{B}T}{\eta_w \eta_o v_w v_o} \exp(\frac{-\lambda^{2}\sigma_{wo}(1+\cos \theta)}{k_B T})
\end{equation}

where $v_w = 0.03 \ \mathrm{nm}^3$ and $v_o = 0.38 \ \mathrm{nm}^3$ denote the molecular volume of water and oil, $h$ the Planck constant and $ \sigma_{wo}(1+\cos \theta)$ corresponds to the reversible work of adhesion between the solid and the liquid, while considering the presence of the other liquid phase. Fig. \ref{DropletParticle} schematically represents the particle diffusing at the water/lipid monolayer/oil interface and the various parameters involved. 

The total friction experienced by the particle diffusing at the lipid monolayer interface is equal to the sum of two uncorrelated terms $\gamma_{tr} = \gamma_{tr,H} + \gamma_{tr,L}$. For the calculation of the hydrodynamic friction $\gamma_{tr,H}$, we approximate the system as a particle dragged along a liquid-liquid interface, with both liquids having identical viscosities, using the result of the hydrodynamic model in \cite{Fischer2006}, the ratio $\eta_o/\eta_w = 1.324$ being close to 1. We set the surface viscosity of the monolayer to $\eta_{m}/2$, where $\eta_m$ is the viscosity of a membrane bilayer with the same lipid composition ($\eta_{m}=2\cdot 10^{-9}\ \mathrm{Pa.s.m}$, see Fig. \ref{FrictionVsPenetrationDepth}). The interfacial tension for similar systems was measured using the pendant drop technique and found to vary between $ 2\text{-}5 \, \mathrm{mN.m^{-1}}$ \cite{Li1995} \cite{Venkatesan2018}. Considering a small contact angle $5^\circ < \theta < 10^\circ$, the characteristic deformation length $\lambda$ ranges between 2.7 and 4.8 nm which can be related to thermally excited capillary waves for low tension interfaces \cite{langevin1992} and is comparable to the length of a lipid molecule. 

This analysis reveals a fundamental difference between the diffusion mechanisms of a particle wrapped by a lipid bilayer, well described by a hydrodynamic model, and a particle diffusing near the water/lipid monolayer/oil interface of a droplet. For the latter system, fluctuations of the three-phase contact line must be considered, as they introduce an additional contribution to the translational drag experienced by the particle diffusing at the interface. The result of our analysis can be used to explain experiments reported in the literature on the slow lateral diffusion of nano- and micro-particles at the oil-water interface, which could not be rationalised so far \cite{Du2012}.

\section{Conclusion}

To conclude, we explored the interaction dynamics between microparticles and giant unilamellar vesicles. The use of DOTAP has already proven effective in driving the wrapping of negatively charged particles by the membrane \cite{Ewins2022}. Here, we demonstrated that this adhesion force can overcome the restoring force of an optical trap as well as the contributions of membrane tension and bending, sometimes leading to vesicle rupture and reaching several nN. We highlighted the impact of penetration depth on translational friction and performed simulations to evaluate the impact of the membrane region wrapping the particle and the water gap existing between the particle and the membrane. We measured that the membrane viscosity is significantly affected by membrane-bound substructures and consequently by an effective lipid layer thickness. The study of particle diffusion near a single lipid monolayer, separating an aqueous phase from an oil phase, however does not translate in a faster diffusion and lower friction with respect to the particle diffusion in the bilayer system. In a monolayer, the particle is in a partial wetting state, and the particle diffusion is also affected by the partial wetting dynamics. In contrast, in the particle-vesicle system, the lipid bilayer is deformed and prevents the particle to protrude in the two bulk phases. These findings enhance our understanding of transport mechanisms and the interactions of micron-sized particles with model cell membranes. 
By investigating systems with controlled physical and chemical parameters, we were able to reproduce mechanisms such as endocytosis and membrane rupture, which are also observed in more complex biological systems. In this sense, quantitative measurements of energetics and dynamics in model soft matter systems may provide new insights useful in the biomedical field.

\section{Materials and methods}

\subsection{Materials}

Chloroform stock solutions of DOPC (1,2-dioleoyl-sn-glycero-3-phosphocholine), DOTAP (1,2-dioleoyl-3-trimethylammonium-propane) and NBD PE (1,2-dioleoyl-sn-glycero-3-phosphoethanolamine-N-(7-nitro-2-1,3-benzoxadiazol-4-yl)) were purchased from Avanti Polar Lipids. DOPC, DOTAP and NBD PE (fluorescently labelled DOPE) were mixed at molar ratio 94:5:1 with a final lipid concentration of 1 g/L. Polystyrene (PS) (radius $a=0.46  \pm 0.01 \,\mathrm{\mu m}$) and melamine formaldehyde (MF) (radius $a=1.25  \pm 0.03 \,\mathrm{\mu m}$)  spherical particles were purchased from microParticles GmbH. Tetraethyl orthosilicate (TEOS, reagent grade), ammonia (analytical grade), potassium chloride (KCl, analytical grade) and ethanol were purchased from Sigma Aldrich. Branched polyethyleneimine (PEI, Lupasol WF, $\text{M}_{\text{w}} = 25\,000 \, \text{g/mol}$ was purchased from BASF.

 \subsection{Synthesis of silica microparticles}

 Silica microspheres (radius $a=0.60  \pm 0.05 \,\mathrm{\mu m}$) were synthesized according to the protocol described by Yu et al \cite{Yu2016}. Briefly, in a 250 mL semibatch chemical reactor, 65 mL of ethanol, 6.75 mL of water and 0.017 g of KCl are mixed at 300 RPM. A solution composed of 33.3 mL of ethanol and 2 g of TEOS is delivered in this solution at a flow rate of 0.2 mL/min using a syringe pump. The temperature is maintained at 30°C through the reaction. For the surface functionalization of silica microparticles with amine groups, 10 mg of the powder silica particles are dispersed in toluene followed by the addition of  60 µL of APTES. The mixture is left under agitation and reflux for 3 hours.  After this step the silica microparticles are collected by centrifugation (5 minutes, 5000 RCF), and washed (3 times) with ethanol and water.

\subsection{Preparation of giant unilamellar vesicles}

Giant unilamellar vesicles formation was achieved by PVA (polyvinyl alcohol) gel-assisted swelling. A 5\% (w/w) PVA solution in ultrapure water is stirred on a hot plate at 80°C for 3 hours. 100 $\mu$L of solution is then spread inside a PTFE (polytetrafluoroethylene) chamber and dried in an oven at 80°C for 35 min. 10 $\mu$L of lipid solution is spread on the PVA gel film and let 30 min under vacuum to ensure evaporation of the chloroform. The dried lipid bilayers are then hydrated by adding 200 $\mu$L of 150 mM sucrose solution and let for 2 hours to allow an optimal swelling of the vesicles. The GUVs suspension is then collected and transferred in 1 mL of a 158 mM glucose solution, leading to slight deflation and sedimentation due to the osmolarity and density mismatch between the inside and the outside of the vesicles. 

\subsection{Zeta potential measurements}

Zeta potential of the particles and vesicles was assessed using a Zetasizer Nano ZS (Malvern Panalytical) and measured to be $\zeta_{\mathrm{SiO\textsubscript{2}}} = -63 \pm 5 \,\mathrm{mV}$, $\zeta_{\mathrm{PS}} = -54 \pm 4 \,\mathrm{mV}$, $\zeta_{\mathrm{SiO\textsubscript{2}-NH\textsubscript{2}}} = -12 \pm 3 \,\mathrm{mV}$ and $\zeta_{\mathrm{MF}} = +25 \pm 6 \,\mathrm{mV}$ for silica, polystyrene, aminated silica and melamine formaldehyde, respectively. For vesicles made of 94 mol\% DOPC – 5 mol\% DOTAP – 1 mol\% NBD PE we found $\zeta_{\mathrm{DOTAP}} = +33 \pm 12 \,\mathrm{mV}$. Measurements were taken at a pH of 5.6, the equilibrium value for pure water exposed to atmospheric air due to dissolved carbon dioxide \cite{seinfeld2016}.

\subsection{Sample cell preparation}

First, a glass coverslip of 0.17 mm thickness is washed through a series of chloroform/acetone/ethanol/ultrapure water baths and dried with compressed air. The coverslip is then immersed in a 2.5 g/L solution of polyethylenimine (PEI) in ultrapure water for 15 minutes, followed by 3 rinses of 2 minutes each in ultrapure water. This step prevents the rupture of positively charged vesicles on the glass coverslip. A self-adhesive silicone isolator of 9 mm diameter and 1.7 mm thickness is then placed on the coverslip to form the sample cell. Finally, the silicone isolator is filled with 100 µL of 158 mM glucose solution, 10 $\mu$L of GUV solution, and 1 $\mu$L of a 1 g/L particle solution. The sample cell is then sealed with another coverslip to maintain constant osmotic pressure and prevent convection effects.

\subsection{Optical tweezers setup}

The experimental setup comprises a 976 nm FBG single mode laser diode with a maximum power output of 300 mW (Thorlabs BL976-SAG300), coupled with a high numerical aperture objective (Nikon Plan Fluor 100X, NA 1.30 Oil, WD 0.20 mm) facilitating optical trapping. A LED light source and a CMOS camera (Hamamatsu Orca-Flash 4.0 C11440) enable sample visualization. Additionally, the system is equipped with a fluorescence module (Thorlabs OTKB-FL) integrated with a mercury lamp (Nikon C-HGFI).

\

\subsection{Preparation of lipid-coated aqueous droplets in oil}

The preparation of lipid-coated aqueous droplets in dodecane (water-in-oil emulsion) relies on the amphiphilic properties of phospholipids which self-assemble into a monolayer at the water/oil interface, with their hydrophilic heads oriented towards the water and their hydrophobic tails oriented towards the oil. First, 400 $\mu$L of a 10 g/L lipid solution in chloroform is injected into a flask and placed on a rotary evaporator for 3 hours. Afterward, 5 mL of dodecane is added to the flask, which is subsequently placed in an ultrasonic bath at 60°C for 30 minutes, followed by incubation in an oven at 60°C for 12 hours to ensure complete dissolution and uniform dispersion of the lipids. Next, 600 $\mu$L of the dodecane/lipid solution is taken and added to 100 $\mu$L of an aqueous solution of silica particles at 0.02 g/L in an Eppendorf tube. The mixture is then vortexed and repeatedly aspirated and dispensed using a micropipette to ensure thorough mixing.

\subsection{Acquisition and tracking}
\label{subsec:AcquisitionTracking}

Videos were acquired at frequencies ranging from 100 to 900 frames per second (FPS) and the tracking of the center of mass was performed using Blender 3.3.0 software. The 3D trajectory of the particle along the vesicle surface was reconstructed from its 2D projection, under the assumption of a spherical vesicle using the method of Shigyou et al. \cite{Shigyou2016}. 
\begin{equation}
\mathrm{MSD} = \left\langle d_{p}^{2}(t,t+\Delta t) \right\rangle = 4D_{tr}\Delta t
\end{equation}

with $d_{p}$ representing the reconstructed displacement of the particle from $t$ to $t+\Delta t$ on the vesicle surface,

\begin{equation}
\begin{split}
d_{p}(t, t + \Delta t) &= R \cos^{-1} \left( 1 - \frac{(x_t - x_{t+\Delta t})^2}{2R^2} \right. \\
&\quad \left. + \frac{(y_t - y_{t+\Delta t})^2 + (z_t - z_{t+\Delta t})^2}{2R^2} \right)
\end{split}
\end{equation}

and the coordinate $z$ at time $t$ calculated considering a spherical vesicle of radius $R$,

\begin{equation}
    z_{t} = \sqrt{R^{2}-(x^{2}_{t}+y^{2}_{t})}
\end{equation}

\begin{acknowledgments}
We acknowledge the support of the Agence Nationale de la Recherche EDEM (ANR-21-CE06-0042), the Doctoral College of Physics and Chemistry-Physics of Strasbourg, and ITI HiFunMat at the University of Strasbourg.
\end{acknowledgments}

\nocite{*}

\bibliography{references}

\providecommand{\noopsort}[1]{}\providecommand{\singleletter}[1]{#1}%
\begin{thebibliography}{45}%
\makeatletter
\providecommand \@ifxundefined [1]{%
 \@ifx{#1\undefined}
}%
\providecommand \@ifnum [1]{%
 \ifnum #1\expandafter \@firstoftwo
 \else \expandafter \@secondoftwo
 \fi
}%
\providecommand \@ifx [1]{%
 \ifx #1\expandafter \@firstoftwo
 \else \expandafter \@secondoftwo
 \fi
}%
\providecommand \natexlab [1]{#1}%
\providecommand \enquote  [1]{``#1''}%
\providecommand \bibnamefont  [1]{#1}%
\providecommand \bibfnamefont [1]{#1}%
\providecommand \citenamefont [1]{#1}%
\providecommand \href@noop [0]{\@secondoftwo}%
\providecommand \href [0]{\begingroup \@sanitize@url \@href}%
\providecommand \@href[1]{\@@startlink{#1}\@@href}%
\providecommand \@@href[1]{\endgroup#1\@@endlink}%
\providecommand \@sanitize@url [0]{\catcode `\\12\catcode `\$12\catcode `\&12\catcode `\#12\catcode `\^12\catcode `\_12\catcode `\%12\relax}%
\providecommand \@@startlink[1]{}%
\providecommand \@@endlink[0]{}%
\providecommand \url  [0]{\begingroup\@sanitize@url \@url }%
\providecommand \@url [1]{\endgroup\@href {#1}{\urlprefix }}%
\providecommand \urlprefix  [0]{URL }%
\providecommand \Eprint [0]{\href }%
\providecommand \doibase [0]{https://doi.org/}%
\providecommand \selectlanguage [0]{\@gobble}%
\providecommand \bibinfo  [0]{\@secondoftwo}%
\providecommand \bibfield  [0]{\@secondoftwo}%
\providecommand \translation [1]{[#1]}%
\providecommand \BibitemOpen [0]{}%
\providecommand \bibitemStop [0]{}%
\providecommand \bibitemNoStop [0]{.\EOS\space}%
\providecommand \EOS [0]{\spacefactor3000\relax}%
\providecommand \BibitemShut  [1]{\csname bibitem#1\endcsname}%
\let\auto@bib@innerbib\@empty
\bibitem [{\citenamefont {Doherty}\ and\ \citenamefont {McMahon}(2009)}]{Doherty2009}%
  \BibitemOpen
  \bibfield  {author} {\bibinfo {author} {\bibfnamefont {G.~J.}\ \bibnamefont {Doherty}}\ and\ \bibinfo {author} {\bibfnamefont {H.~T.}\ \bibnamefont {McMahon}},\ }\bibfield  {title} {\bibinfo {title} {Mechanisms of endocytosis},\ }\href {https://doi.org/10.1146/annurev.biochem.78.081307.110540} {\bibfield  {journal} {\bibinfo  {journal} {Annual Review of Biochemistry}\ }\textbf {\bibinfo {volume} {78}},\ \bibinfo {pages} {857–902} (\bibinfo {year} {2009})}\BibitemShut {NoStop}%
\bibitem [{\citenamefont {Mellman}(1996)}]{Mellman1996}%
  \BibitemOpen
  \bibfield  {author} {\bibinfo {author} {\bibfnamefont {I.}~\bibnamefont {Mellman}},\ }\bibfield  {title} {\bibinfo {title} {Endocytosis and molecular sorting},\ }\href {https://doi.org/10.1146/annurev.cellbio.12.1.575} {\bibfield  {journal} {\bibinfo  {journal} {Annual Review of Cell and Developmental Biology}\ }\textbf {\bibinfo {volume} {12}},\ \bibinfo {pages} {575–625} (\bibinfo {year} {1996})}\BibitemShut {NoStop}%
\bibitem [{\citenamefont {Yong}\ \emph {et~al.}(2020)\citenamefont {Yong}, \citenamefont {Valiyaveettil},\ and\ \citenamefont {Tang}}]{Yong2020}%
  \BibitemOpen
  \bibfield  {author} {\bibinfo {author} {\bibfnamefont {C.}~\bibnamefont {Yong}}, \bibinfo {author} {\bibfnamefont {S.}~\bibnamefont {Valiyaveettil}},\ and\ \bibinfo {author} {\bibfnamefont {B.}~\bibnamefont {Tang}},\ }\bibfield  {title} {\bibinfo {title} {Toxicity of microplastics and nanoplastics in mammalian systems},\ }\href {https://doi.org/10.3390/ijerph17051509} {\bibfield  {journal} {\bibinfo  {journal} {International Journal of Environmental Research and Public Health}\ }\textbf {\bibinfo {volume} {17}},\ \bibinfo {pages} {1509} (\bibinfo {year} {2020})}\BibitemShut {NoStop}%
\bibitem [{\citenamefont {Agudo-Canalejo}\ and\ \citenamefont {Lipowsky}(2019)}]{AgudoCanalejo2019}%
  \BibitemOpen
  \bibfield  {author} {\bibinfo {author} {\bibfnamefont {J.}~\bibnamefont {Agudo-Canalejo}}\ and\ \bibinfo {author} {\bibfnamefont {R.}~\bibnamefont {Lipowsky}},\ }\bibfield  {title} {\bibinfo {title} {Particle–membrane interactions},\ }in\ \href {https://doi.org/10.1201/9781315152516} {\emph {\bibinfo {booktitle} {The Giant Vesicle Book}}},\ \bibinfo {editor} {edited by\ \bibinfo {editor} {\bibfnamefont {R.}~\bibnamefont {Dimova}}\ and\ \bibinfo {editor} {\bibfnamefont {C.}~\bibnamefont {Marques}}}\ (\bibinfo  {publisher} {CRC Press},\ \bibinfo {year} {2019})\BibitemShut {NoStop}%
\bibitem [{\citenamefont {Israelachvili}(2011)}]{Israelachvili2011}%
  \BibitemOpen
  \bibfield  {author} {\bibinfo {author} {\bibfnamefont {J.~N.}\ \bibnamefont {Israelachvili}},\ }\bibinfo {title} {Interactions of biological membranes and structures},\ in\ \href {https://doi.org/10.1016/b978-0-12-391927-4.10021-0} {\emph {\bibinfo {booktitle} {Intermolecular and Surface Forces}}}\ (\bibinfo  {publisher} {Elsevier},\ \bibinfo {year} {2011})\ p.\ \bibinfo {pages} {577–616}\BibitemShut {NoStop}%
\bibitem [{\citenamefont {van~der Wel}\ \emph {et~al.}(2017)\citenamefont {van~der Wel}, \citenamefont {Heinrich},\ and\ \citenamefont {Kraft}}]{vanderWel2017}%
  \BibitemOpen
  \bibfield  {author} {\bibinfo {author} {\bibfnamefont {C.}~\bibnamefont {van~der Wel}}, \bibinfo {author} {\bibfnamefont {D.}~\bibnamefont {Heinrich}},\ and\ \bibinfo {author} {\bibfnamefont {D.~J.}\ \bibnamefont {Kraft}},\ }\bibfield  {title} {\bibinfo {title} {Microparticle assembly pathways on lipid membranes},\ }\href {https://doi.org/10.1016/j.bpj.2017.07.019} {\bibfield  {journal} {\bibinfo  {journal} {Biophysical Journal}\ }\textbf {\bibinfo {volume} {113}},\ \bibinfo {pages} {1037–1046} (\bibinfo {year} {2017})}\BibitemShut {NoStop}%
\bibitem [{\citenamefont {Spanke}\ \emph {et~al.}(2020)\citenamefont {Spanke}, \citenamefont {Style}, \citenamefont {Fran\c{c}ois-Martin}, \citenamefont {Feofilova}, \citenamefont {Eisentraut}, \citenamefont {Kress}, \citenamefont {Agudo-Canalejo},\ and\ \citenamefont {Dufresne}}]{Spanke2020}%
  \BibitemOpen
  \bibfield  {author} {\bibinfo {author} {\bibfnamefont {H.~T.}\ \bibnamefont {Spanke}}, \bibinfo {author} {\bibfnamefont {R.~W.}\ \bibnamefont {Style}}, \bibinfo {author} {\bibfnamefont {C.}~\bibnamefont {Fran\c{c}ois-Martin}}, \bibinfo {author} {\bibfnamefont {M.}~\bibnamefont {Feofilova}}, \bibinfo {author} {\bibfnamefont {M.}~\bibnamefont {Eisentraut}}, \bibinfo {author} {\bibfnamefont {H.}~\bibnamefont {Kress}}, \bibinfo {author} {\bibfnamefont {J.}~\bibnamefont {Agudo-Canalejo}},\ and\ \bibinfo {author} {\bibfnamefont {E.~R.}\ \bibnamefont {Dufresne}},\ }\bibfield  {title} {\bibinfo {title} {Wrapping of microparticles by floppy lipid vesicles},\ }\bibfield  {journal} {\bibinfo  {journal} {Physical Review Letters}\ }\textbf {\bibinfo {volume} {125}},\ \href {https://doi.org/10.1103/physrevlett.125.198102} {10.1103/physrevlett.125.198102} (\bibinfo {year} {2020})\BibitemShut {NoStop}%
\bibitem [{\citenamefont {Ewins}\ \emph {et~al.}(2022)\citenamefont {Ewins}, \citenamefont {Han}, \citenamefont {Bharti}, \citenamefont {Robinson}, \citenamefont {Velev},\ and\ \citenamefont {Dimova}}]{Ewins2022}%
  \BibitemOpen
  \bibfield  {author} {\bibinfo {author} {\bibfnamefont {E.~J.}\ \bibnamefont {Ewins}}, \bibinfo {author} {\bibfnamefont {K.}~\bibnamefont {Han}}, \bibinfo {author} {\bibfnamefont {B.}~\bibnamefont {Bharti}}, \bibinfo {author} {\bibfnamefont {T.}~\bibnamefont {Robinson}}, \bibinfo {author} {\bibfnamefont {O.~D.}\ \bibnamefont {Velev}},\ and\ \bibinfo {author} {\bibfnamefont {R.}~\bibnamefont {Dimova}},\ }\bibfield  {title} {\bibinfo {title} {Controlled adhesion, membrane pinning and vesicle transport by janus particles},\ }\href {https://doi.org/10.1039/d1cc07026f} {\bibfield  {journal} {\bibinfo  {journal} {Chemical Communications}\ }\textbf {\bibinfo {volume} {58}},\ \bibinfo {pages} {3055–3058} (\bibinfo {year} {2022})}\BibitemShut {NoStop}%
\bibitem [{\citenamefont {Fery}\ \emph {et~al.}(2003)\citenamefont {Fery}, \citenamefont {Moya}, \citenamefont {Puech}, \citenamefont {Brochard-Wyart},\ and\ \citenamefont {Mohwald}}]{Fery2003}%
  \BibitemOpen
  \bibfield  {author} {\bibinfo {author} {\bibfnamefont {A.}~\bibnamefont {Fery}}, \bibinfo {author} {\bibfnamefont {S.}~\bibnamefont {Moya}}, \bibinfo {author} {\bibfnamefont {P.-H.}\ \bibnamefont {Puech}}, \bibinfo {author} {\bibfnamefont {F.}~\bibnamefont {Brochard-Wyart}},\ and\ \bibinfo {author} {\bibfnamefont {H.}~\bibnamefont {Mohwald}},\ }\bibfield  {title} {\bibinfo {title} {Interaction of polyelectrolyte coated beads with phospholipid vesicles},\ }\href {https://doi.org/10.1016/s1631-0705(03)00030-6} {\bibfield  {journal} {\bibinfo  {journal} {Comptes Rendus. Physique}\ }\textbf {\bibinfo {volume} {4}},\ \bibinfo {pages} {259–264} (\bibinfo {year} {2003})}\BibitemShut {NoStop}%
\bibitem [{\citenamefont {Dietrich}\ \emph {et~al.}(1997)\citenamefont {Dietrich}, \citenamefont {Angelova},\ and\ \citenamefont {Pouligny}}]{Dietrich1997}%
  \BibitemOpen
  \bibfield  {author} {\bibinfo {author} {\bibfnamefont {C.}~\bibnamefont {Dietrich}}, \bibinfo {author} {\bibfnamefont {M.}~\bibnamefont {Angelova}},\ and\ \bibinfo {author} {\bibfnamefont {B.}~\bibnamefont {Pouligny}},\ }\bibfield  {title} {\bibinfo {title} {Adhesion of latex spheres to giant phospholipid vesicles: Statics and dynamics},\ }\href {https://doi.org/10.1051/jp2:1997208} {\bibfield  {journal} {\bibinfo  {journal} {Journal de Physique II}\ }\textbf {\bibinfo {volume} {7}},\ \bibinfo {pages} {1651–1682} (\bibinfo {year} {1997})}\BibitemShut {NoStop}%
\bibitem [{\citenamefont {Dimova}\ \emph {et~al.}(1999)\citenamefont {Dimova}, \citenamefont {Dietrich}, \citenamefont {Hadjiisky}, \citenamefont {Danov},\ and\ \citenamefont {Pouligny}}]{Dimova1999}%
  \BibitemOpen
  \bibfield  {author} {\bibinfo {author} {\bibfnamefont {R.}~\bibnamefont {Dimova}}, \bibinfo {author} {\bibfnamefont {C.}~\bibnamefont {Dietrich}}, \bibinfo {author} {\bibfnamefont {A.}~\bibnamefont {Hadjiisky}}, \bibinfo {author} {\bibfnamefont {K.}~\bibnamefont {Danov}},\ and\ \bibinfo {author} {\bibfnamefont {B.}~\bibnamefont {Pouligny}},\ }\bibfield  {title} {\bibinfo {title} {Falling ball viscosimetry of giant vesicle membranes: Finite-size effects},\ }\href {https://doi.org/10.1007/s100510051042} {\bibfield  {journal} {\bibinfo  {journal} {The European Physical Journal B}\ }\textbf {\bibinfo {volume} {12}},\ \bibinfo {pages} {589–598} (\bibinfo {year} {1999})}\BibitemShut {NoStop}%
\bibitem [{\citenamefont {Shigyou}\ \emph {et~al.}(2016)\citenamefont {Shigyou}, \citenamefont {Nagai},\ and\ \citenamefont {Hamada}}]{Shigyou2016}%
  \BibitemOpen
  \bibfield  {author} {\bibinfo {author} {\bibfnamefont {K.}~\bibnamefont {Shigyou}}, \bibinfo {author} {\bibfnamefont {K.~H.}\ \bibnamefont {Nagai}},\ and\ \bibinfo {author} {\bibfnamefont {T.}~\bibnamefont {Hamada}},\ }\bibfield  {title} {\bibinfo {title} {Lateral diffusion of a submicrometer particle on a lipid bilayer membrane},\ }\href {https://doi.org/10.1021/acs.langmuir.6b02448} {\bibfield  {journal} {\bibinfo  {journal} {Langmuir}\ }\textbf {\bibinfo {volume} {32}},\ \bibinfo {pages} {13771–13777} (\bibinfo {year} {2016})}\BibitemShut {NoStop}%
\bibitem [{\citenamefont {Bahrami}\ \emph {et~al.}(2014)\citenamefont {Bahrami}, \citenamefont {Raatz}, \citenamefont {Agudo-Canalejo}, \citenamefont {Michel}, \citenamefont {Curtis}, \citenamefont {Hall}, \citenamefont {Gradzielski}, \citenamefont {Lipowsky},\ and\ \citenamefont {Weikl}}]{Bahrami2014}%
  \BibitemOpen
  \bibfield  {author} {\bibinfo {author} {\bibfnamefont {A.~H.}\ \bibnamefont {Bahrami}}, \bibinfo {author} {\bibfnamefont {M.}~\bibnamefont {Raatz}}, \bibinfo {author} {\bibfnamefont {J.}~\bibnamefont {Agudo-Canalejo}}, \bibinfo {author} {\bibfnamefont {R.}~\bibnamefont {Michel}}, \bibinfo {author} {\bibfnamefont {E.~M.}\ \bibnamefont {Curtis}}, \bibinfo {author} {\bibfnamefont {C.~K.}\ \bibnamefont {Hall}}, \bibinfo {author} {\bibfnamefont {M.}~\bibnamefont {Gradzielski}}, \bibinfo {author} {\bibfnamefont {R.}~\bibnamefont {Lipowsky}},\ and\ \bibinfo {author} {\bibfnamefont {T.~R.}\ \bibnamefont {Weikl}},\ }\bibfield  {title} {\bibinfo {title} {Wrapping of nanoparticles by membranes},\ }\href {https://doi.org/10.1016/j.cis.2014.02.012} {\bibfield  {journal} {\bibinfo  {journal} {Advances in Colloid and Interface Science}\ }\textbf {\bibinfo {volume} {208}},\ \bibinfo {pages} {214–224} (\bibinfo {year} {2014})}\BibitemShut {NoStop}%
\bibitem [{\citenamefont {Agudo-Canalejo}\ and\ \citenamefont {Lipowsky}(2015)}]{AgudoCanalejo2015}%
  \BibitemOpen
  \bibfield  {author} {\bibinfo {author} {\bibfnamefont {J.}~\bibnamefont {Agudo-Canalejo}}\ and\ \bibinfo {author} {\bibfnamefont {R.}~\bibnamefont {Lipowsky}},\ }\bibfield  {title} {\bibinfo {title} {Critical particle sizes for the engulfment of nanoparticles by membranes and vesicles with bilayer asymmetry},\ }\href {https://doi.org/10.1021/acsnano.5b01285} {\bibfield  {journal} {\bibinfo  {journal} {ACS Nano}\ }\textbf {\bibinfo {volume} {9}},\ \bibinfo {pages} {3704–3720} (\bibinfo {year} {2015})}\BibitemShut {NoStop}%
\bibitem [{\citenamefont {Saffman}\ and\ \citenamefont {Delbr\"{u}ck}(1975)}]{Saffman1975}%
  \BibitemOpen
  \bibfield  {author} {\bibinfo {author} {\bibfnamefont {P.~G.}\ \bibnamefont {Saffman}}\ and\ \bibinfo {author} {\bibfnamefont {M.}~\bibnamefont {Delbr\"{u}ck}},\ }\bibfield  {title} {\bibinfo {title} {Brownian motion in biological membranes.},\ }\href {https://doi.org/10.1073/pnas.72.8.3111} {\bibfield  {journal} {\bibinfo  {journal} {Proceedings of the National Academy of Sciences}\ }\textbf {\bibinfo {volume} {72}},\ \bibinfo {pages} {3111–3113} (\bibinfo {year} {1975})}\BibitemShut {NoStop}%
\bibitem [{\citenamefont {Hughes}\ \emph {et~al.}(1981)\citenamefont {Hughes}, \citenamefont {Pailthorpe},\ and\ \citenamefont {White}}]{Hughes1981}%
  \BibitemOpen
  \bibfield  {author} {\bibinfo {author} {\bibfnamefont {B.~D.}\ \bibnamefont {Hughes}}, \bibinfo {author} {\bibfnamefont {B.~A.}\ \bibnamefont {Pailthorpe}},\ and\ \bibinfo {author} {\bibfnamefont {L.~R.}\ \bibnamefont {White}},\ }\bibfield  {title} {\bibinfo {title} {The translational and rotational drag on a cylinder moving in a membrane},\ }\href {https://doi.org/10.1017/s0022112081000785} {\bibfield  {journal} {\bibinfo  {journal} {Journal of Fluid Mechanics}\ }\textbf {\bibinfo {volume} {110}},\ \bibinfo {pages} {349–372} (\bibinfo {year} {1981})}\BibitemShut {NoStop}%
\bibitem [{\citenamefont {Evans}\ and\ \citenamefont {Sackmann}(1988)}]{Evans1988}%
  \BibitemOpen
  \bibfield  {author} {\bibinfo {author} {\bibfnamefont {E.}~\bibnamefont {Evans}}\ and\ \bibinfo {author} {\bibfnamefont {E.}~\bibnamefont {Sackmann}},\ }\bibfield  {title} {\bibinfo {title} {Translational and rotational drag coefficients for a disk moving in a liquid membrane associated with a rigid substrate},\ }\href {https://doi.org/10.1017/s0022112088003106} {\bibfield  {journal} {\bibinfo  {journal} {Journal of Fluid Mechanics}\ }\textbf {\bibinfo {volume} {194}},\ \bibinfo {pages} {553} (\bibinfo {year} {1988})}\BibitemShut {NoStop}%
\bibitem [{\citenamefont {Danov}\ \emph {et~al.}(1995)\citenamefont {Danov}, \citenamefont {Aust}, \citenamefont {Durst},\ and\ \citenamefont {Lange}}]{Danov1995}%
  \BibitemOpen
  \bibfield  {author} {\bibinfo {author} {\bibfnamefont {K.}~\bibnamefont {Danov}}, \bibinfo {author} {\bibfnamefont {R.}~\bibnamefont {Aust}}, \bibinfo {author} {\bibfnamefont {F.}~\bibnamefont {Durst}},\ and\ \bibinfo {author} {\bibfnamefont {U.}~\bibnamefont {Lange}},\ }\bibfield  {title} {\bibinfo {title} {Influence of the surface viscosity on the hydrodynamic resistance and surface diffusivity of a large brownian particle},\ }\href {https://doi.org/10.1006/jcis.1995.1426} {\bibfield  {journal} {\bibinfo  {journal} {Journal of Colloid and Interface Science}\ }\textbf {\bibinfo {volume} {175}},\ \bibinfo {pages} {36–45} (\bibinfo {year} {1995})}\BibitemShut {NoStop}%
\bibitem [{\citenamefont {Danov}\ \emph {et~al.}(2000)\citenamefont {Danov}, \citenamefont {Dimova},\ and\ \citenamefont {Pouligny}}]{Danov2000}%
  \BibitemOpen
  \bibfield  {author} {\bibinfo {author} {\bibfnamefont {K.~D.}\ \bibnamefont {Danov}}, \bibinfo {author} {\bibfnamefont {R.}~\bibnamefont {Dimova}},\ and\ \bibinfo {author} {\bibfnamefont {B.}~\bibnamefont {Pouligny}},\ }\bibfield  {title} {\bibinfo {title} {Viscous drag of a solid sphere straddling a spherical or flat surface},\ }\href {https://doi.org/10.1063/1.1289692} {\bibfield  {journal} {\bibinfo  {journal} {Physics of Fluids}\ }\textbf {\bibinfo {volume} {12}},\ \bibinfo {pages} {2711–2722} (\bibinfo {year} {2000})}\BibitemShut {NoStop}%
\bibitem [{\citenamefont {Fischer}\ \emph {et~al.}(2006)\citenamefont {Fischer}, \citenamefont {Dhar},\ and\ \citenamefont {Heinig}}]{Fischer2006}%
  \BibitemOpen
  \bibfield  {author} {\bibinfo {author} {\bibfnamefont {T.~M.}\ \bibnamefont {Fischer}}, \bibinfo {author} {\bibfnamefont {P.}~\bibnamefont {Dhar}},\ and\ \bibinfo {author} {\bibfnamefont {P.}~\bibnamefont {Heinig}},\ }\bibfield  {title} {\bibinfo {title} {The viscous drag of spheres and filaments moving in membranes or monolayers},\ }\href {https://doi.org/10.1017/s002211200600022x} {\bibfield  {journal} {\bibinfo  {journal} {Journal of Fluid Mechanics}\ }\textbf {\bibinfo {volume} {558}},\ \bibinfo {pages} {451} (\bibinfo {year} {2006})}\BibitemShut {NoStop}%
\bibitem [{\citenamefont {Lipowsky}(2013)}]{Lipowsky2013}%
  \BibitemOpen
  \bibfield  {author} {\bibinfo {author} {\bibfnamefont {R.}~\bibnamefont {Lipowsky}},\ }\bibfield  {title} {\bibinfo {title} {Spontaneous tubulation of membranes and vesicles reveals membrane tension generated by spontaneous curvature},\ }\href {https://doi.org/10.1039/c2fd20105d} {\bibfield  {journal} {\bibinfo  {journal} {Faraday Discuss.}\ }\textbf {\bibinfo {volume} {161}},\ \bibinfo {pages} {305–331} (\bibinfo {year} {2013})}\BibitemShut {NoStop}%
\bibitem [{\citenamefont {Spanke}\ \emph {et~al.}(2022)\citenamefont {Spanke}, \citenamefont {Agudo-Canalejo}, \citenamefont {Tran}, \citenamefont {Style},\ and\ \citenamefont {Dufresne}}]{Spanke2022}%
  \BibitemOpen
  \bibfield  {author} {\bibinfo {author} {\bibfnamefont {H.~T.}\ \bibnamefont {Spanke}}, \bibinfo {author} {\bibfnamefont {J.}~\bibnamefont {Agudo-Canalejo}}, \bibinfo {author} {\bibfnamefont {D.}~\bibnamefont {Tran}}, \bibinfo {author} {\bibfnamefont {R.~W.}\ \bibnamefont {Style}},\ and\ \bibinfo {author} {\bibfnamefont {E.~R.}\ \bibnamefont {Dufresne}},\ }\bibfield  {title} {\bibinfo {title} {Dynamics of spontaneous wrapping of microparticles by floppy lipid membranes},\ }\bibfield  {journal} {\bibinfo  {journal} {Physical Review Research}\ }\textbf {\bibinfo {volume} {4}},\ \href {https://doi.org/10.1103/physrevresearch.4.023080} {10.1103/physrevresearch.4.023080} (\bibinfo {year} {2022})\BibitemShut {NoStop}%
\bibitem [{\citenamefont {Deserno}(2004)}]{Deserno2004}%
  \BibitemOpen
  \bibfield  {author} {\bibinfo {author} {\bibfnamefont {M.}~\bibnamefont {Deserno}},\ }\bibfield  {title} {\bibinfo {title} {Elastic deformation of a fluid membrane upon colloid binding},\ }\bibfield  {journal} {\bibinfo  {journal} {Physical Review E}\ }\textbf {\bibinfo {volume} {69}},\ \href {https://doi.org/10.1103/physreve.69.031903} {10.1103/physreve.69.031903} (\bibinfo {year} {2004})\BibitemShut {NoStop}%
\bibitem [{\citenamefont {Fessler}\ \emph {et~al.}(2023)\citenamefont {Fessler}, \citenamefont {Sharma}, \citenamefont {Muller},\ and\ \citenamefont {Stocco}}]{Fessler2023}%
  \BibitemOpen
  \bibfield  {author} {\bibinfo {author} {\bibfnamefont {F.}~\bibnamefont {Fessler}}, \bibinfo {author} {\bibfnamefont {V.}~\bibnamefont {Sharma}}, \bibinfo {author} {\bibfnamefont {P.}~\bibnamefont {Muller}},\ and\ \bibinfo {author} {\bibfnamefont {A.}~\bibnamefont {Stocco}},\ }\bibfield  {title} {\bibinfo {title} {Entry of microparticles into giant lipid vesicles by optical tweezers},\ }\bibfield  {journal} {\bibinfo  {journal} {Physical Review E}\ }\textbf {\bibinfo {volume} {107}},\ \href {https://doi.org/10.1103/physreve.107.l052601} {10.1103/physreve.107.l052601} (\bibinfo {year} {2023})\BibitemShut {NoStop}%
\bibitem [{\citenamefont {Lipowsky}(2019)}]{Lipowsky2019}%
  \BibitemOpen
  \bibfield  {author} {\bibinfo {author} {\bibfnamefont {R.}~\bibnamefont {Lipowsky}},\ }\bibfield  {title} {\bibinfo {title} {Understanding giant vesicles: A theoretical perspective},\ }in\ \href {https://doi.org/10.1201/9781315152516} {\emph {\bibinfo {booktitle} {The Giant Vesicle Book}}},\ \bibinfo {editor} {edited by\ \bibinfo {editor} {\bibfnamefont {R.}~\bibnamefont {Dimova}}\ and\ \bibinfo {editor} {\bibfnamefont {C.}~\bibnamefont {Marques}}}\ (\bibinfo  {publisher} {CRC Press},\ \bibinfo {year} {2019})\BibitemShut {NoStop}%
\bibitem [{\citenamefont {Olbrich}\ \emph {et~al.}(2000)\citenamefont {Olbrich}, \citenamefont {Rawicz}, \citenamefont {Needham},\ and\ \citenamefont {Evans}}]{Olbrich2000}%
  \BibitemOpen
  \bibfield  {author} {\bibinfo {author} {\bibfnamefont {K.}~\bibnamefont {Olbrich}}, \bibinfo {author} {\bibfnamefont {W.}~\bibnamefont {Rawicz}}, \bibinfo {author} {\bibfnamefont {D.}~\bibnamefont {Needham}},\ and\ \bibinfo {author} {\bibfnamefont {E.}~\bibnamefont {Evans}},\ }\bibfield  {title} {\bibinfo {title} {Water permeability and mechanical strength of polyunsaturated lipid bilayers},\ }\href {https://doi.org/10.1016/s0006-3495(00)76294-1} {\bibfield  {journal} {\bibinfo  {journal} {Biophysical Journal}\ }\textbf {\bibinfo {volume} {79}},\ \bibinfo {pages} {321–327} (\bibinfo {year} {2000})}\BibitemShut {NoStop}%
\bibitem [{\citenamefont {Moy}\ \emph {et~al.}(1999)\citenamefont {Moy}, \citenamefont {Jiao}, \citenamefont {Hillmann}, \citenamefont {Lehmann},\ and\ \citenamefont {Sano}}]{Moy1999}%
  \BibitemOpen
  \bibfield  {author} {\bibinfo {author} {\bibfnamefont {V.~T.}\ \bibnamefont {Moy}}, \bibinfo {author} {\bibfnamefont {Y.}~\bibnamefont {Jiao}}, \bibinfo {author} {\bibfnamefont {T.}~\bibnamefont {Hillmann}}, \bibinfo {author} {\bibfnamefont {H.}~\bibnamefont {Lehmann}},\ and\ \bibinfo {author} {\bibfnamefont {T.}~\bibnamefont {Sano}},\ }\bibfield  {title} {\bibinfo {title} {Adhesion energy of receptor-mediated interaction measured by elastic deformation},\ }\href {https://doi.org/10.1016/s0006-3495(99)77322-4} {\bibfield  {journal} {\bibinfo  {journal} {Biophysical Journal}\ }\textbf {\bibinfo {volume} {76}},\ \bibinfo {pages} {1632–1638} (\bibinfo {year} {1999})}\BibitemShut {NoStop}%
\bibitem [{\citenamefont {Faxén}(1922)}]{Faxn1922}%
  \BibitemOpen
  \bibfield  {author} {\bibinfo {author} {\bibfnamefont {H.}~\bibnamefont {Faxén}},\ }\bibfield  {title} {\bibinfo {title} {Der widerstand gegen die bewegung einer starren kugel in einer z\"{a}hen fl\"{u}ssigkeit, die zwischen zwei parallelen ebenen w\"{a}nden eingeschlossen ist},\ }\href {https://doi.org/10.1002/andp.19223731003} {\bibfield  {journal} {\bibinfo  {journal} {Annalen der Physik}\ }\textbf {\bibinfo {volume} {373}},\ \bibinfo {pages} {89–119} (\bibinfo {year} {1922})}\BibitemShut {NoStop}%
\bibitem [{\citenamefont {R\"{a}dler}\ \emph {et~al.}(1995)\citenamefont {R\"{a}dler}, \citenamefont {Feder}, \citenamefont {Strey},\ and\ \citenamefont {Sackmann}}]{Rdler1995}%
  \BibitemOpen
  \bibfield  {author} {\bibinfo {author} {\bibfnamefont {J.~O.}\ \bibnamefont {R\"{a}dler}}, \bibinfo {author} {\bibfnamefont {T.~J.}\ \bibnamefont {Feder}}, \bibinfo {author} {\bibfnamefont {H.~H.}\ \bibnamefont {Strey}},\ and\ \bibinfo {author} {\bibfnamefont {E.}~\bibnamefont {Sackmann}},\ }\bibfield  {title} {\bibinfo {title} {Fluctuation analysis of tension-controlled undulation forces between giant vesicles and solid substrates},\ }\href {https://doi.org/10.1103/physreve.51.4526} {\bibfield  {journal} {\bibinfo  {journal} {Physical Review E}\ }\textbf {\bibinfo {volume} {51}},\ \bibinfo {pages} {4526–4536} (\bibinfo {year} {1995})}\BibitemShut {NoStop}%
\bibitem [{\citenamefont {Johnson}\ \emph {et~al.}(1991)\citenamefont {Johnson}, \citenamefont {Bayerl}, \citenamefont {McDermott}, \citenamefont {Adam}, \citenamefont {Rennie}, \citenamefont {Thomas},\ and\ \citenamefont {Sackmann}}]{Johnson1991}%
  \BibitemOpen
  \bibfield  {author} {\bibinfo {author} {\bibfnamefont {S.}~\bibnamefont {Johnson}}, \bibinfo {author} {\bibfnamefont {T.}~\bibnamefont {Bayerl}}, \bibinfo {author} {\bibfnamefont {D.}~\bibnamefont {McDermott}}, \bibinfo {author} {\bibfnamefont {G.}~\bibnamefont {Adam}}, \bibinfo {author} {\bibfnamefont {A.}~\bibnamefont {Rennie}}, \bibinfo {author} {\bibfnamefont {R.}~\bibnamefont {Thomas}},\ and\ \bibinfo {author} {\bibfnamefont {E.}~\bibnamefont {Sackmann}},\ }\bibfield  {title} {\bibinfo {title} {Structure of an adsorbed dimyristoylphosphatidylcholine bilayer measured with specular reflection of neutrons},\ }\href {https://doi.org/10.1016/s0006-3495(91)82222-6} {\bibfield  {journal} {\bibinfo  {journal} {Biophysical Journal}\ }\textbf {\bibinfo {volume} {59}},\ \bibinfo {pages} {289–294} (\bibinfo {year} {1991})}\BibitemShut {NoStop}%
\bibitem [{\citenamefont {Bayerl}\ and\ \citenamefont {Bloom}(1990)}]{Bayerl1990}%
  \BibitemOpen
  \bibfield  {author} {\bibinfo {author} {\bibfnamefont {T.}~\bibnamefont {Bayerl}}\ and\ \bibinfo {author} {\bibfnamefont {M.}~\bibnamefont {Bloom}},\ }\bibfield  {title} {\bibinfo {title} {Physical properties of single phospholipid bilayers adsorbed to micro glass beads. a new vesicular model system studied by 2h-nuclear magnetic resonance},\ }\href {https://doi.org/10.1016/s0006-3495(90)82382-1} {\bibfield  {journal} {\bibinfo  {journal} {Biophysical Journal}\ }\textbf {\bibinfo {volume} {58}},\ \bibinfo {pages} {357–362} (\bibinfo {year} {1990})}\BibitemShut {NoStop}%
\bibitem [{\citenamefont {Faizi}\ \emph {et~al.}(2022)\citenamefont {Faizi}, \citenamefont {Dimova},\ and\ \citenamefont {Vlahovska}}]{Faizi2022}%
  \BibitemOpen
  \bibfield  {author} {\bibinfo {author} {\bibfnamefont {H.~A.}\ \bibnamefont {Faizi}}, \bibinfo {author} {\bibfnamefont {R.}~\bibnamefont {Dimova}},\ and\ \bibinfo {author} {\bibfnamefont {P.~M.}\ \bibnamefont {Vlahovska}},\ }\bibfield  {title} {\bibinfo {title} {A vesicle microrheometer for high-throughput viscosity measurements of lipid and polymer membranes},\ }\href {https://doi.org/10.1016/j.bpj.2022.02.015} {\bibfield  {journal} {\bibinfo  {journal} {Biophysical Journal}\ }\textbf {\bibinfo {volume} {121}},\ \bibinfo {pages} {910–918} (\bibinfo {year} {2022})}\BibitemShut {NoStop}%
\bibitem [{\citenamefont {Hormel}\ \emph {et~al.}(2014)\citenamefont {Hormel}, \citenamefont {Kurihara}, \citenamefont {Brennan}, \citenamefont {Wozniak},\ and\ \citenamefont {Parthasarathy}}]{Hormel2014}%
  \BibitemOpen
  \bibfield  {author} {\bibinfo {author} {\bibfnamefont {T.~T.}\ \bibnamefont {Hormel}}, \bibinfo {author} {\bibfnamefont {S.~Q.}\ \bibnamefont {Kurihara}}, \bibinfo {author} {\bibfnamefont {M.~K.}\ \bibnamefont {Brennan}}, \bibinfo {author} {\bibfnamefont {M.~C.}\ \bibnamefont {Wozniak}},\ and\ \bibinfo {author} {\bibfnamefont {R.}~\bibnamefont {Parthasarathy}},\ }\bibfield  {title} {\bibinfo {title} {Measuring lipid membrane viscosity using rotational and translational probe diffusion},\ }\bibfield  {journal} {\bibinfo  {journal} {Physical Review Letters}\ }\textbf {\bibinfo {volume} {112}},\ \href {https://doi.org/10.1103/physrevlett.112.188101} {10.1103/physrevlett.112.188101} (\bibinfo {year} {2014})\BibitemShut {NoStop}%
\bibitem [{\citenamefont {Iwahashi}\ \emph {et~al.}(1990)\citenamefont {Iwahashi}, \citenamefont {Yamaguchi}, \citenamefont {Ogura},\ and\ \citenamefont {Suzuki}}]{Iwahashi1990}%
  \BibitemOpen
  \bibfield  {author} {\bibinfo {author} {\bibfnamefont {M.}~\bibnamefont {Iwahashi}}, \bibinfo {author} {\bibfnamefont {Y.}~\bibnamefont {Yamaguchi}}, \bibinfo {author} {\bibfnamefont {Y.}~\bibnamefont {Ogura}},\ and\ \bibinfo {author} {\bibfnamefont {M.}~\bibnamefont {Suzuki}},\ }\bibfield  {title} {\bibinfo {title} {Dynamical structures of normal alkanes, alcohols, and fatty acids in the liquid state as determined by viscosity, self-diffusion coefficient, infrared spectra, and 13cnmr spin-lattice relaxation time measurements},\ }\href {https://doi.org/10.1246/bcsj.63.2154} {\bibfield  {journal} {\bibinfo  {journal} {Bulletin of the Chemical Society of Japan}\ }\textbf {\bibinfo {volume} {63}},\ \bibinfo {pages} {2154–2158} (\bibinfo {year} {1990})}\BibitemShut {NoStop}%
\bibitem [{\citenamefont {Horton}\ \emph {et~al.}(2010)\citenamefont {Horton}, \citenamefont {H\"{o}fling}, \citenamefont {R\"{a}dler},\ and\ \citenamefont {Franosch}}]{Horton2010}%
  \BibitemOpen
  \bibfield  {author} {\bibinfo {author} {\bibfnamefont {M.~R.}\ \bibnamefont {Horton}}, \bibinfo {author} {\bibfnamefont {F.}~\bibnamefont {H\"{o}fling}}, \bibinfo {author} {\bibfnamefont {J.~O.}\ \bibnamefont {R\"{a}dler}},\ and\ \bibinfo {author} {\bibfnamefont {T.}~\bibnamefont {Franosch}},\ }\bibfield  {title} {\bibinfo {title} {Development of anomalous diffusion among crowding proteins},\ }\href {https://doi.org/10.1039/b924149c} {\bibfield  {journal} {\bibinfo  {journal} {Soft Matter}\ }\textbf {\bibinfo {volume} {6}},\ \bibinfo {pages} {2648} (\bibinfo {year} {2010})}\BibitemShut {NoStop}%
\bibitem [{\citenamefont {Skaug}\ \emph {et~al.}(2011)\citenamefont {Skaug}, \citenamefont {Faller},\ and\ \citenamefont {Longo}}]{Skaug2011}%
  \BibitemOpen
  \bibfield  {author} {\bibinfo {author} {\bibfnamefont {M.~J.}\ \bibnamefont {Skaug}}, \bibinfo {author} {\bibfnamefont {R.}~\bibnamefont {Faller}},\ and\ \bibinfo {author} {\bibfnamefont {M.~L.}\ \bibnamefont {Longo}},\ }\bibfield  {title} {\bibinfo {title} {Correlating anomalous diffusion with lipid bilayer membrane structure using single molecule tracking and atomic force microscopy},\ }\bibfield  {journal} {\bibinfo  {journal} {The Journal of Chemical Physics}\ }\textbf {\bibinfo {volume} {134}},\ \href {https://doi.org/10.1063/1.3596377} {10.1063/1.3596377} (\bibinfo {year} {2011})\BibitemShut {NoStop}%
\bibitem [{\citenamefont {Daddi-Moussa-Ider}\ \emph {et~al.}(2016)\citenamefont {Daddi-Moussa-Ider}, \citenamefont {Guckenberger},\ and\ \citenamefont {Gekle}}]{DaddiMoussaIder2016}%
  \BibitemOpen
  \bibfield  {author} {\bibinfo {author} {\bibfnamefont {A.}~\bibnamefont {Daddi-Moussa-Ider}}, \bibinfo {author} {\bibfnamefont {A.}~\bibnamefont {Guckenberger}},\ and\ \bibinfo {author} {\bibfnamefont {S.}~\bibnamefont {Gekle}},\ }\bibfield  {title} {\bibinfo {title} {Long-lived anomalous thermal diffusion induced by elastic cell membranes on nearby particles},\ }\bibfield  {journal} {\bibinfo  {journal} {Physical Review E}\ }\textbf {\bibinfo {volume} {93}},\ \href {https://doi.org/10.1103/physreve.93.012612} {10.1103/physreve.93.012612} (\bibinfo {year} {2016})\BibitemShut {NoStop}%
\bibitem [{\citenamefont {Boniello}\ \emph {et~al.}(2015)\citenamefont {Boniello}, \citenamefont {Blanc}, \citenamefont {Fedorenko}, \citenamefont {Medfai}, \citenamefont {Mbarek}, \citenamefont {In}, \citenamefont {Gross}, \citenamefont {Stocco},\ and\ \citenamefont {Nobili}}]{Boniello2015}%
  \BibitemOpen
  \bibfield  {author} {\bibinfo {author} {\bibfnamefont {G.}~\bibnamefont {Boniello}}, \bibinfo {author} {\bibfnamefont {C.}~\bibnamefont {Blanc}}, \bibinfo {author} {\bibfnamefont {D.}~\bibnamefont {Fedorenko}}, \bibinfo {author} {\bibfnamefont {M.}~\bibnamefont {Medfai}}, \bibinfo {author} {\bibfnamefont {N.~B.}\ \bibnamefont {Mbarek}}, \bibinfo {author} {\bibfnamefont {M.}~\bibnamefont {In}}, \bibinfo {author} {\bibfnamefont {M.}~\bibnamefont {Gross}}, \bibinfo {author} {\bibfnamefont {A.}~\bibnamefont {Stocco}},\ and\ \bibinfo {author} {\bibfnamefont {M.}~\bibnamefont {Nobili}},\ }\bibfield  {title} {\bibinfo {title} {Brownian diffusion of a partially wetted colloid},\ }\href {https://doi.org/10.1038/nmat4348} {\bibfield  {journal} {\bibinfo  {journal} {Nature Materials}\ }\textbf {\bibinfo {volume} {14}},\ \bibinfo {pages} {908–911} (\bibinfo {year} {2015})}\BibitemShut {NoStop}%
\bibitem [{\citenamefont {Blake}(1993)}]{blake1993dynamic}%
  \BibitemOpen
  \bibfield  {author} {\bibinfo {author} {\bibfnamefont {T.~D.}\ \bibnamefont {Blake}},\ }\bibfield  {title} {\bibinfo {title} {Dynamic contact angles and wetting kinetics},\ }in\ \href@noop {} {\emph {\bibinfo {booktitle} {Wettability}}},\ \bibinfo {editor} {edited by\ \bibinfo {editor} {\bibfnamefont {J.}~\bibnamefont {Berg}}}\ (\bibinfo  {publisher} {Marcel Dekker Inc.},\ \bibinfo {address} {New York},\ \bibinfo {year} {1993})\ p.\ \bibinfo {pages} {251}\BibitemShut {NoStop}%
\bibitem [{\citenamefont {Li}\ \emph {et~al.}(1995)\citenamefont {Li}, \citenamefont {Miller}, \citenamefont {W\"{u}stneck}, \citenamefont {M\"{o}hwald},\ and\ \citenamefont {Neumann}}]{Li1995}%
  \BibitemOpen
  \bibfield  {author} {\bibinfo {author} {\bibfnamefont {J.}~\bibnamefont {Li}}, \bibinfo {author} {\bibfnamefont {R.}~\bibnamefont {Miller}}, \bibinfo {author} {\bibfnamefont {R.}~\bibnamefont {W\"{u}stneck}}, \bibinfo {author} {\bibfnamefont {H.}~\bibnamefont {M\"{o}hwald}},\ and\ \bibinfo {author} {\bibfnamefont {A.}~\bibnamefont {Neumann}},\ }\bibfield  {title} {\bibinfo {title} {Use of pendent drop technique as a film balance at liquid/liquid interfaces},\ }\href {https://doi.org/10.1016/0927-7757(94)03062-5} {\bibfield  {journal} {\bibinfo  {journal} {Colloids and Surfaces A: Physicochemical and Engineering Aspects}\ }\textbf {\bibinfo {volume} {96}},\ \bibinfo {pages} {295–299} (\bibinfo {year} {1995})}\BibitemShut {NoStop}%
\bibitem [{\citenamefont {Venkatesan}\ \emph {et~al.}(2018)\citenamefont {Venkatesan}, \citenamefont {Taylor}, \citenamefont {Basham}, \citenamefont {Brady}, \citenamefont {Collier},\ and\ \citenamefont {Sarles}}]{Venkatesan2018}%
  \BibitemOpen
  \bibfield  {author} {\bibinfo {author} {\bibfnamefont {G.~A.}\ \bibnamefont {Venkatesan}}, \bibinfo {author} {\bibfnamefont {G.~J.}\ \bibnamefont {Taylor}}, \bibinfo {author} {\bibfnamefont {C.~M.}\ \bibnamefont {Basham}}, \bibinfo {author} {\bibfnamefont {N.~G.}\ \bibnamefont {Brady}}, \bibinfo {author} {\bibfnamefont {C.~P.}\ \bibnamefont {Collier}},\ and\ \bibinfo {author} {\bibfnamefont {S.~A.}\ \bibnamefont {Sarles}},\ }\bibfield  {title} {\bibinfo {title} {Evaporation-induced monolayer compression improves droplet interface bilayer formation using unsaturated lipids},\ }\bibfield  {journal} {\bibinfo  {journal} {Biomicrofluidics}\ }\textbf {\bibinfo {volume} {12}},\ \href {https://doi.org/10.1063/1.5016523} {10.1063/1.5016523} (\bibinfo {year} {2018})\BibitemShut {NoStop}%
\bibitem [{\citenamefont {Langevin}(1992)}]{langevin1992}%
  \BibitemOpen
  \bibinfo {editor} {\bibfnamefont {D.}~\bibnamefont {Langevin}},\ ed.,\ \href@noop {} {\emph {\bibinfo {title} {Light Scattering by Liquid Surfaces and Complementary Techniques}}}\ (\bibinfo  {publisher} {Marcel Dekker Inc.},\ \bibinfo {address} {New York},\ \bibinfo {year} {1992})\BibitemShut {NoStop}%
\bibitem [{\citenamefont {Du}\ \emph {et~al.}(2012)\citenamefont {Du}, \citenamefont {Liddle},\ and\ \citenamefont {Berglund}}]{Du2012}%
  \BibitemOpen
  \bibfield  {author} {\bibinfo {author} {\bibfnamefont {K.}~\bibnamefont {Du}}, \bibinfo {author} {\bibfnamefont {J.~A.}\ \bibnamefont {Liddle}},\ and\ \bibinfo {author} {\bibfnamefont {A.~J.}\ \bibnamefont {Berglund}},\ }\bibfield  {title} {\bibinfo {title} {Three-dimensional real-time tracking of nanoparticles at an oil–water interface},\ }\href {https://doi.org/10.1021/la300292r} {\bibfield  {journal} {\bibinfo  {journal} {Langmuir}\ }\textbf {\bibinfo {volume} {28}},\ \bibinfo {pages} {9181–9188} (\bibinfo {year} {2012})}\BibitemShut {NoStop}%
\bibitem [{\citenamefont {Yu}\ \emph {et~al.}(2016)\citenamefont {Yu}, \citenamefont {Cong}, \citenamefont {Xue}, \citenamefont {Tian}, \citenamefont {Xu}, \citenamefont {Peng},\ and\ \citenamefont {Yang}}]{Yu2016}%
  \BibitemOpen
  \bibfield  {author} {\bibinfo {author} {\bibfnamefont {B.}~\bibnamefont {Yu}}, \bibinfo {author} {\bibfnamefont {H.}~\bibnamefont {Cong}}, \bibinfo {author} {\bibfnamefont {L.}~\bibnamefont {Xue}}, \bibinfo {author} {\bibfnamefont {C.}~\bibnamefont {Tian}}, \bibinfo {author} {\bibfnamefont {X.}~\bibnamefont {Xu}}, \bibinfo {author} {\bibfnamefont {Q.}~\bibnamefont {Peng}},\ and\ \bibinfo {author} {\bibfnamefont {S.}~\bibnamefont {Yang}},\ }\bibfield  {title} {\bibinfo {title} {Synthesis and modification of monodisperse silica microspheres for uplc separation of c60 and c70},\ }\href {https://doi.org/10.1039/c5ay02655e} {\bibfield  {journal} {\bibinfo  {journal} {Analytical Methods}\ }\textbf {\bibinfo {volume} {8}},\ \bibinfo {pages} {919–924} (\bibinfo {year} {2016})}\BibitemShut {NoStop}%
\bibitem [{\citenamefont {Seinfeld}\ and\ \citenamefont {Pandis}(2016)}]{seinfeld2016}%
  \BibitemOpen
  \bibfield  {author} {\bibinfo {author} {\bibfnamefont {J.~H.}\ \bibnamefont {Seinfeld}}\ and\ \bibinfo {author} {\bibfnamefont {S.~N.}\ \bibnamefont {Pandis}},\ }\href@noop {} {\emph {\bibinfo {title} {Atmospheric Chemistry and Physics: From Air Pollution to Climate Change}}},\ \bibinfo {edition} {3rd}\ ed.\ (\bibinfo  {publisher} {Wiley},\ \bibinfo {year} {2016})\BibitemShut {NoStop}%
\end{thebibliography}%

\end{document}